\documentclass{aastex702}
\usepackage[T1]{fontenc}

\newcommand{\PWVunits}{kg/m$^{2}$}

\newcommand{\PWVprecision}{0.4} 
\newcommand{\PWVaccuracy}{1} 

\newcommand{\biasgLABERA}{-0.18} 
\newcommand{\springPWVmean}{3.0} 
\newcommand{\summerPWVmean}{5.4} 
\newcommand{\RgLABERA}{0.89} 
\newcommand{\PWVRMSEgLABERA}{1.1} 
\newcommand{\PWVstdgLABERA}{1.1} 

\newcommand{\SeptemberAmplitude}{1.2} 
\newcommand{\NovemberAmplitude}{1.7}
\newcommand{\FebruaryAmplitude}{3.8} 
\newcommand{\springRwind}{0.69} 
\newcommand{\summerRwind}{0.81}  
\newcommand{\meanSpringAmplitude}{1.1} 
\newcommand{\meanSummerAmplitude}{2.7} 

\begin{document}

\title{Measuring Precipitable Water Vapor through Dual-Band GPS at Rubin Observatory}

\author[0009-0007-7841-1052]{Anja~Weyant}
\affiliation{
    Pittsburgh Particle Physics, Astrophysics, and Cosmology Center (PITT PACC).
    Physics and Astronomy Department, University of Pittsburgh,
    Pittsburgh, PA 15260, USA
}
\email{anw421@pitt.edu}

\author[0000-0001-7113-1233]{W.~M.~Wood-Vasey}
\affiliation{
    Pittsburgh Particle Physics, Astrophysics, and Cosmology Center (PITT PACC).
    Physics and Astronomy Department, University of Pittsburgh,
    Pittsburgh, PA 15260, USA
}
\email{wmwv@pitt.edu}

\author[0000-0002-3205-2484]{Elana~K.~Urbach}
\affiliation{
    Department of Physics, Harvard University,
    17 Oxford Street, Cambridge, MA, USA
}
\affiliation{
    Vera C. Rubin Observatory, La Serena, Chile
}
\email{eurbach@g.harvard.edu}

\author[]{Sara~Gelven}
\affiliation{
    Pittsburgh Particle Physics, Astrophysics, and Cosmology Center (PITT PACC).
    Physics and Astronomy Department, University of Pittsburgh,
    Pittsburgh, PA 15260, USA
}
\email{seg191@pitt.edu}

\author[]{Marcos~Tamargo-Arizmedi}
\affiliation{
    Pittsburgh Particle Physics, Astrophysics, and Cosmology Center (PITT PACC).
    Physics and Astronomy Department, University of Pittsburgh,
    Pittsburgh, PA 15260, USA
}
\email{met228@pitt.edu}

\begin{abstract}

  Precipitable water vapor (PWV) in Earth's atmosphere is a known source of systematic error in ground-based astronomical measurements, particularly at wavelengths $\lambda>7,000$~\AA. PWV produces wavelength-dependent extinction in the spectral energy distributions of observed sources.  For high-precision applications, such as supernova cosmology, the transmission function must be measured accurately and precisely for each observation. The Vera C. Rubin Observatory is located in the Chilean Andes, a region known for dry conditions. However, even at this location, time-variable absorption by water vapor is a factor. Signal delays from Global Navigation Satellite Systems (GNSS) enable a robust, easy, 24-hour monitoring of PWV. In this paper we present measurements from the first six months of GNSS-PWV at Rubin Observatory.  We verified our analysis by replicating results from the International GNSS Service (IGS), demonstrating high consistency.  We then installed and commissioned a Trimble Net R9 receiver at Allegheny Observatory for a commissioning run before it was relocated to Rubin Observatory. At Rubin Observatory we have been continuously measuring PWV since 2025 September 5.  We find average PWV levels of \springPWVmean{}~\PWVunits{} in Southern Hemisphere spring (September to November) and \summerPWVmean{}~\PWVunits{} in summer (December to February).  The PWV at Rubin also has a clear daily variation with an amplitude of \SeptemberAmplitude--\FebruaryAmplitude~\PWVunits{} that is highly correlated with the east-west wind strength.  We conclude that the proposed GNSS-based processing method achieves an accuracy of \PWVaccuracy{}~\PWVunits{} with \PWVprecision{}~\PWVunits{} precision, confirming the capability for continuous, high-temporal-resolution PWV monitoring to support demanding LSST science cases.

\end{abstract}

\section{Introduction}

The Vera C. Rubin Observatory, through its Legacy Survey of Space and Time (LSST), is an ambitious wide-field ground-based survey with four key science goals: exploring dark energy and dark matter, cataloging the solar system, furthering time-domain Astronomy, and mapping the Milky Way \citep{LSST19}.  The resulting catalog will be thousands of times larger than any other previously assembled. The ambitious scientific goals of the LSST demand equally ambitious photometric calibration by requiring a 5~mmag photometric precision and 10~mmag absolute calibration \citep{LSST19}.  For ground-based observations this is only achievable by accounting for effects caused by Earth's atmosphere.  Accounting for absorption due to water vapor is a challenging one because it has a complex structure with wavelength.

The precipitable water vapor (PWV) is measured in \PWVunits{} or equivalently mm\footnote{Liquid water has a density of 1,000 kg/m$^{3}$, 1 \PWVunits{} equals 1 mm of water depth. To avoid confusion with mm of distance, we use \PWVunits{}.} and is the total amount of water in the atmosphere condensed. The altitude of the observing site therefore determines the upper bound of PWV - a high-altitude site simply has less atmosphere above it. The other main contributors to PWV are temperature - warm air can hold more water vapor - and large-scale atmospheric circulation which transports moist air masses \citep{Trenberth05}.
While we care about the water vapor along a line of sight,
PWV is traditionally calculated as the equivalent amount of water vapor at zenith\footnote{This is analogous to astronomical observations being corrected to an airmass of 1, although there are some geometric subtleties about the height distribution of water vapor vs other atmospheric constituents of absorption and scattering.}.
PWV absorption of light significantly affects the atmospheric transmission function in the redder optical and near-infrared bands \citep{Wood-Vasey22, Burke14}.
Astronomical images are traditionally calibrated using a reference catalog of stars to model corrections for color, airmass, and a higher order color-airmass term.  This approach works well when the difference in spectral energy distribution (SED) between the target and reference stars is well described by a smooth monotonic function. It also adequately accounts for effects that are smoother with wavelength such as Mie scattering due to aerosols in the atmosphere.

However, this approach is no longer sufficient in an era of high-precision photometry \citep[e.g.]{Burke18}.
Objects like Type Ia supernovae (SNe~Ia) exhibit a spectral energy distribution that is feature rich, unlike the smooth Rayleigh-Jeans tail of typical stars.
The measured brightness of targets will vary as a function of redshift as different spectral features move across PWV absorption bands. A misestimate in PWV of just a few \PWVunits{} (a few mm) results in a bias of up to 0.02~mag in distance modulus for low redshift SN (z < 0.2) and using stars to calibrate SNe~Ia fails to fully capture the complex structure of PWV \citep{Wood-Vasey22}.
SNe~Ia derive their cosmological utility from their standard nature and calibration errors like this are a leading source of uncertainty, accounting for up to 50\% of the total error in cosmological parameters \citep{Scolnic18}.
A detailed understanding of the atmosphere over the course of a night is essential to providing real-time photometric corrections to the observations.

To this end, LSST will utilize the Forward Global Calibration Model (FGCM) for photometric calibration \citep{Burke18}.  This method takes as input measurements of the optical properties of the instrument and (when available) atmospheric monitoring data from external equipment. These are combined with the repeated broad-band imaging data of the survey and models of the instrument and atmosphere to create a forward model that jointly estimates the spatial- and time-dependence of the passbands, including atmospheric terms when auxiliary data is sparse or absent. The passband for any observation can be combined with an estimate of source spectrum to yield a chromatically-corrected magnitude, allowing for sub-percent calibration. Knowing the atmospheric transmission function for each image is key to calibrating SNe~Ia for cosmology. It is particularly important for objects with a different SED than the stars that will dominate the calibration solutions. GPS-derived precipitable water vapor is one such source of external atmospheric monitoring data, and can be incorporated into FGCM's forward model in place of, or as a check on, its self-calibrated PWV estimate.

There are multiple constellations of Global Navigation Satellite Systems (GNSS): GPS, GLONASS, Galileo, and BeiDou.  While they differ in details of orbits, frequencies, and accuracy, they operate based on the same physical principles of using the time delays in the radio signals from orbiting atomic clocks enable the determination of position based on measurements from several satellites simultaneously.  Signals from GNSS can also be used to determine PWV by measuring the additional delay of radio signals between satellites and a ground-based receiver with a known (or static) position.  The delay combined with surface temperature and pressure measurements can be used to measure the amount of PWV in the atmosphere.  The combined measurements from satellites across the sky is expressed in a single number, the equivalent PWV at zenith.  This technique is well-established in the atmospheric sciences community\footnote{http://www.suominet.ucar.edu/} and is increasingly being adopted in observational astronomy \citep{Blake11,Querel14,Baker17,Burke18,Li18,Wood-Vasey22}.

\subsection{How PWV Affects GNSS signals}

The effect of the atmosphere on GNSS signals is dominated by two main components: the troposphere and the ionosphere. Fortunately the effect of the ionosphere is frequency dependent and can be 99.9\% removed using a dual-frequency receiver \citep[see, e.g.,][]{brunner91}.  The refractivity $N$ of the troposphere is divided into a hydrostatic component and a wet component $N = N_{\text{hydro}} + N_{\text{wet}}$ \citep{Davis85}. The hydrostatic component is a result of dry gases in the atmosphere, namely $N_2$ and $O_2$ (but can include a contribution from water vapor due to the non-dipole component of water vapor refractivity \citep{Bevis92}).  Its effect can be accurately modeled using ideal gas laws and depends on temperature and atmospheric pressure \citep{Saastamoinen72}.   This term is approximately a 1--2~m delay in the zenith direction.  The wet component is caused solely by water vapor and condensed water in the form of clouds. This component depends on the weather conditions and can vary in an unpredictable way. As a result, for high accuracy positioning, the delay caused by the wet component is estimated together with the coordinates of the receiver. This delay is around tens of centimeters in the zenith direction \citep{Sanz13}.

\subsection{Fundamentals of GNSS Measurements}

GNSS satellites transmit in the L band (1--2~GHz) at two or more frequencies; for GPS these are L1 (1575.42~MHz), L2 (1227.60~MHz), and L5 (1176.45~MHz). There are three main types of signals.  The {\em ranging code} is a binary code unique to each satellite. This signal allows one to measure the travel time between the receiver and the satellite. The other two signals are the carrier phase and the navigation data.  The {\em carrier phase} is a sinusoidal signal transmitted at a given frequency and the {\em navigation data} is a binary-coded message that transmits among other things, the satellite position, clock information, and satellite health.

To achieve high-accuracy positioning, and therefore measurements of the tropospheric delay, the carrier phase must be used in addition to the code signal.  The carrier phase measurements are very precise, typically two orders of magnitude more precise than the code at the level of a few millimeters.  However, they are not accurate.  They contain unknown {\em ambiguities} because the receiver cannot determine the number of complete wavelengths between itself and the satellite at the start of tracking. When the receiver subsequently loses lock on the signal (a cycle slip), the ambiguity changes by an unknown integer amount. Resolving these ambiguities together with other parameters in the navigation solution is what allows us to make accurate estimates of the signal delay through the troposphere and thus measure PWV.

We refer the reader to \cite{Sanz13} for a detailed explanation of GNSS positioning. Here we briefly summarize the key equations.  The main observable in GNSS positioning is the travel time of a signal to propagate from the satellite antenna to the receiver antenna.  This travel time multiplied by the speed of light gives the {\em pseudorange}. This distance is not equal to the geometric distance due to synchronization errors between receiver and satellite clocks, signal propagation effects due to the troposphere and ionosphere, instrumental delays, multipath delays (when a signal reaches the receiver via multiple paths due to reflection), and receiver noise.  More precisely, the ionosphere-free combination of frequencies $f_1$ and $f_2$ for the pseudorange $R$ and carrier phase $\Phi$ can be written as
\begin{equation}
  R = \frac{f_{1}^{2} \cdot R_{1} - f_{2}^{2} \cdot R_{2}}{f_{1}^{2} - f_{2}^{2}}= \rho + c(\delta t_{\text {rcv}}-\delta t^{\text {sat}})+{\text{Tr}} + M + \epsilon
  \label{eq:code}
\end{equation}
\begin{equation}
  \Phi = \frac{f_{1}^{2} \cdot \Phi_{1} - f_{2}^{2} \cdot \Phi_{2}}{f_{1}^{2} - f_{2}^{2}}=\rho + c(\delta t_{\text {rcv}}-\delta t^{\text {sat}})+{\text{Tr}} + B + \lambda_{N}\omega + m + \epsilon.
  \label{eq:carrierphase}
\end{equation}
where $\rho$ is the geometric distance, $\text{Tr}$ is the delay caused by the troposphere,  $M$ and $m$ are the multipath for the pseudorange and carrier phase, and $\epsilon$ is the receiver noise. 
$\delta t_{\text {rcv}}$ and $\delta t^{\text {sat}}$ are the receiver and satellite clock offsets from GNSS time and include relativistic satellite clock corrections. These clocks are defined in such a way that the inter-frequency bias between code instrument delays at $f_1$ and $f_2$ cancel in the ionosphere-free combination of code measurements. This bias is commonly called Differential Code Bias (DCB) and is instead included in the bias term $B$ in Eq.~\ref{eq:carrierphase}. $B$ also includes the ambiguities for each carrier phase.
The instrument delays, multipath, and receiver noise are code type and frequency dependent and will therefore be different for each satellite and between GNSS constellations.
Lastly,  $\lambda_{N}\omega$ is called the {\em windup} and is due to the circular polarization of the carrier phase signal. The measured phase depends on the relative orientation of the satellite and receiver antennas. As the satellite rotates to point its solar panels toward the Sun, the rotation causes a phase variation that can be misinterpreted as a range change.

{\em Precise Point Positioning} (PPP) is a technique that can achieve milimeter-centimeter level accuracy in position using code and carrier phase observations from a single dual-frequency GNSS receiver \citep{Zumberge97,Kouba01,Gao02,Gao04}. Indeed PPP is so successful that it has applications in geodetic studies such as monitoring tectonic plate motion, earthquake deformation, hydrologic loading applications such as quantifying absolute sea level rise, and volcanic inflation and deflation \citep[see, e.g.][and references therein]{Hydrogeodesy}. This technique estimates $\text{Tr}$ in addition to the carrier phase bias, and the position and clock offset of the receiver using the ionosphere-free combinations of frequencies. It relies on the availability of precise satellite orbit and clock products such as those provided by the International GNSS Service (IGS) \citep{IGS}. The technique also requires accurate modeling to the centimeter level of measurement effects due to, e.g., tidal motions, the satellite and receiver antenna phase centers, and relativistic effects.

Eq.~\ref{eq:code} and Eq.~\ref{eq:carrierphase} for each satellite in view at a single epoch comprise the foundation of the navigation equations.  They form a set that can be linearized about the approximate receiver coordinates. A minimum of five satellites are needed to solve for the position $(x,y,z)$, clock offsets, the tropospheric delay and to resolve the carrier phase bias for each satellite. A Kalman filter algorithm is employed to continually update these parameters.  Kalman filtering is a recursive algorithm that processes a series of noisy measurements typically by finding a weighted least squares solution of a linearized system of equations \citep{Kalman60}. The algorithm continually updates the parameters based on its previous estimates. As a result, a convergence time is required for high accuracy measurements.

SuomiNet \citep{Ware00}, an early GPS PWV network used by astronomers, was designed for near-real-time delivery to support meteorological research, which led to processing choices suited to low latency, including hourly solutions stacked over a sliding 24-hour window and \emph{predicted} satellite orbit products \citep{Rocken97, Rocken02}. As initially documented, SuomiNet also solved all receivers jointly as a network. This is in contrast to PPP, which processes a receiver independently, and our implementation uses longer data arcs (3 days) with IGS final products, the most accurate available. PPP has been shown to produce fewer outliers than network solutions in tropospheric delay estimates from which PWV is derived \citep{Stepniak22}. Because these strategies differ substantially, we consider it worthwhile to revisit GNSS-derived PWV for astronomical calibration using the PPP strategy.

The goal of this paper is to present the software and configuration, and to outline the process for making these measurements at Rubin Observatory.  We present the software in Section~\ref{sec:glab} and outline how to retrieve PWV in Section~\ref{sec:PWVretrieval}. We describe auxiliary data used in this paper, the receiver, initial commissioning at Allegheny Observatory, and installation at Rubin Observatory in Section~\ref{sec:data}.  We preform a detailed analysis of the software and processing strategy in Section~\ref{sec:analysis}.  We present PWV values from the first six months of operation at Rubin is presented in Section~\ref{sec:results}.  We end the paper with how to access the data (Section~\ref{sec:dataaccess}), future work (Section~\ref{sec:futurework}), and finally summarize our results in Section~\ref{sec:summary}.

\section{gLAB}\label{sec:glab}

We processed the GNSS observations using the GNSS-Lab tool suite \citep[gLAB v6.0.0;][]{gLAB}\footnote{\url{https://gage.upc.edu/en/learning-materials/software-tools/glab-tool-suite}}
which is an open source\footnote{Released under the Apache License, Version 2.0} GNSS analysis software developed by Catalonia Technical University and the European Space Agency.  gLAB is one of the best performing stand alone post processing software packages available for free download \citep{Mendez18, Vazquez23}.
The software and supporting documentation are an excellent resource for understanding GNSS architecture and processing strategies and we encourage the interested reader to peruse them.

In PPP mode, gLAB performs model corrections for many effects necessary to achieve cm-level accuracy.  It corrects for the satellite clock offset using external precise orbits and clocks that are published with a lag time of a few weeks.  It also considers the satellite movement during flight time, Earth's rotation during the signal flight time, and corrects for the satellite mass center to antenna phase center. That is, gLAB further accounts for the difference between the base of the antenna (from which position is usually referenced) to the antenna phase center.  An external ANTEX file\footnote{\url{https://igs.org/wg/antenna/\#files}} is required to perform the phase center corrections for satellite and receiver antennas. gLAB also performs the ``wind up'' correction for the carrier phase and applies relativistic clock and path range corrections.

Ambiguity resolution plays an important role in high-precision applications.  Once the ambiguities in the carrier phase measurements are fixed, the carrier phase becomes an unambiguous pseudorange and can be accurate at the mm level. gLAB employs a solution procedure called {\em floating ambiguities}.  Even though the carrier phase ambiguity is nominally integer cycle slips, the ambiguities are estimated by the Kalman filter as real numbers, not integers. This is in part to account for DCBs. The carrier phase bias is considered ``constant'' until a cycle slip occurs, and then estimated as white noise \citep{Sanz13}. In this implementation, a long convergence time is necessary to fix the ambiguities ($\approx 1$~hr+). We use the default configuration in gLAB PPP mode for cycle-slip detection and estimation.

The signal delay depends on the path through the troposphere and thus depends on the satellite elevation angle. The tropospheric delay in gLAB considers different obliquity factors, or mapping functions, for the wet and dry components and is modeled as
\begin{equation}
        \text{Tr}(E) = \text{Tr}_{z,{\text{dry}}} \cdot M_{\text{dry}}(E) + \text{Tr}_{z,{\text{wet}}}  \cdot M_{\text{wet}}(E)\label{eq:tropglab}
\end{equation}
where $\text{Tr}_{z,{\text{dry}}} $ is the vertical delay from the dry component, $\text{Tr}_{z,{\text{wet}}}$ is the vertical delay from the wet component, $ M_{\text{dry}}(E)$ and $M_{\text{wet}}(E)$  are the mapping functions for the wet and dry components as a function of elevation angle $E$. The dry component is modeled as

\begin{equation}
        \text{Tr}_{z,{\text{dry}}} = \alpha e^{-\beta H}\label{eq:dry}
\end{equation}
where $\alpha = 2.3$~m, $\beta = 0.116 \cdot 10^{-3}$~m$^{-1}$ and $H$ is the height above sea level.  The vertical wet component is modeled as

\begin{equation}
        \text{Tr}_{z,{\text{wet}}} = \text{Tr}_{z0,{\text{wet}}} + \Delta \text{Tr}_{z,{\text{wet}}}\label{eq:slantdelay}
\end{equation}
where $\text{Tr}_{z0,{\text{wet}}} = 0.1$~m and $\Delta \text{Tr}_{z,{\text{wet}}}$, is estimated as a random walk process in the navigation Kalman filter with position coordinates and other parameters.  In gLAB the accuracy of  $\Delta \text{Tr}_{z,{\text{wet}}}$ in static positioning is at the cm level.  The vertical delays are combined with Niell mapping functions, one for the hydrostatic component and one for the wet component \citep{Niell96}, which converts zenith path delays to line-of-sight delays. The Niell mapping functions do not require surface meteorological data, only the receiver latitude and height and the day of year.  In this implementation the wet delay is modeled as a single zenith value that is isotropic, i.e. the same in all azimuths at a given elevation. Other PPP software further decomposes the troposphere into horizontal gradient terms that quantify how much the delay departs from azimuthal symmetry about the receiver (see, e.g., \citet{Chen97}).

In static PPP mode the receiver clock is taken as a white-noise process and the receiver coordinates are taken as constants. In our gLAB processing we additionally adjust the elevation mask to discard satellites low on the horizon which experience increased noise due to the ionosphere and multipath reflections. We also discard satellites under eclipse conditions and satellites whose health flag is set to ``unhealthy'' or ``marginal'' in the broadcast navigation message.
A summary of the general processing strategy is presented in Table~\ref{table:glabconfiguration}.

\begin{center}
\begin{deluxetable}{ll}
\tabletypesize{\footnotesize}
\tablewidth{0pt}
\tablecolumns{2}
\tablecaption{gLAB Configuration}
\label{table:glabconfiguration}
\tablehead{\colhead{Modelling Option} &
        \colhead{Setting}}
\startdata
Satellite Orbits & Precise (IGS Final) \\
Clock Products &  IGS Final \\
Phase Center Offsets & igs20.atx \\
Sampling Rate & 300s \\
GNSS Type & GPS  \\
Elevation Mask & 7$^{\circ}$ \\
Troposphere Model & Simple Nominal $+$ Niell Mapping\\
Ionosphere correction & Ionosphere-free linear combination\\
A priori receiver position & SINEX \\
Observation processed & Code and phase \\
\enddata
\end{deluxetable}
\end{center}

\section{PWV Retrieval}\label{sec:PWVretrieval}

Converting the vertical or zenith tropospheric delay (ZTD) to PWV requires two steps; first we subtract the zenith hydrostatic delay and then we scale the remaining wet delay to PWV by a multiplicative factor. The ZTD can be decomposed into the zenith hydrostatic delay (ZHD) and the zenith wet delay (ZWD) as follows
\begin{equation}
  \text{ZTD} = \text{ZHD} + \text{ZWD}.
\end{equation}
We calculate the ZHD according to the Saastamoinen model \citep{Saastamoinen72}\footnote{gLAB is modeling the slant delay in Eq.~\ref{eq:tropglab} as a dry component plus a wet component.  We are taking the ZTD as measured by gLAB, the sum of both vertical delays, and subtracting a better model for the hydrostatic component, rather than using Eq.~\ref{eq:dry}. This accounts for any mismodeling of the hydrostatic delay in gLAB.}. This model is considered one of the best performing models for the hydrostatic delay \citep{Mendes99, Tuka13} providing sub-millimeter accuracy. This model relies on an assumption of hydrostatic equilibrium and requires accurate pressure measurements at the location of the receiver. We note that this model was derived fitting to data across the entire globe and while it performs well overall, it may not model some variations locally.  There is evidence in some locations that the RMS of the Saastamoinen model may be as large as a few millimeters in ZHD and additionally exhibit seasonal variability \citep{Zhang16, Feng20, Sam23}.
For reference, a misestimate of $+2$~mm in the hydrostatic model translates to about $-0.3$~\PWVunits{} in the inferred PWV.
The Saastamoinen model is given by \citep{Davis85, Abbasy17}
\begin{equation}\label{eq:saas}
  \text{ZHD} = \frac{0.0022767 \cdot {P_s}}{1 - 0.00266 \cdot \cos (2 \phi) - 0.00028 \cdot H}
\end{equation}
where $P_s$ is the surface pressure (hPa), and $\phi$ and $H$ are the geodetic latitude and height (km) of the receiver.
For our GNSS antenna at Rubin at a latitude of -30:14:41, elevation of H=2.687~km, and for a typical pressure of 742 hPa, a characteristic value of ZHD would be 1.69~m.

\begin{figure}
  \centering
  \includegraphics[width=0.6\textwidth]{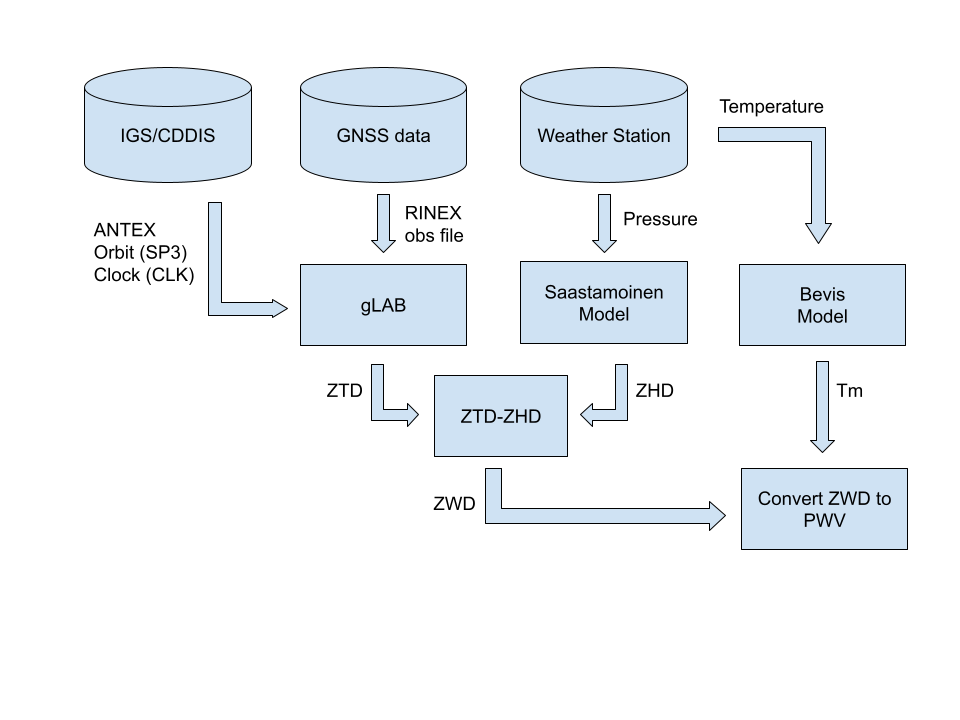}
  \caption{Flow chart describing the process to obtain PWV from GNSS data.  The RINEX observation files (Section \ref{sec:appendix}) along with precise orbit (SP3) and clock files (CLK) from IGS, archived by CDDIS (see Section \ref{sec:IGS}), are inputs into the gLAB software. An additional ANTEX file is required to correct for antenna phase center offsets and variations. The main output of interest from gLAB is zenith total delay (ZTD). The zenith wet delay (ZWD) is obtained by subtracting the hydrostatic component (ZHD) from the ZTD. ZHD is derived according to the Saastamoinen model \citep{Saastamoinen72} using pressure measured at the Rubin weather station, corrected for the height difference between the weather station and the antenna.  Finally, the ZWD is converted to PWV using the conversion factor $\Pi$, which depends on the weighted mean temperature of the atmosphere $T_m$. $T_m$ in this work is approximated by the relationship presented in \citet{Bevis94} which takes surface temperature as input.}
  \label{fig:flowchart}
\end{figure}

After isolating the ZWD term, PWV is found through a conversion factor $\Pi$ given by \citep{Bevis94}:
\begin{equation}
  \text{PWV} = \Pi \cdot \text{ZWD},
\end{equation}
\begin{equation}
  \Pi = \frac{10^6}{\rho R_\nu[(\kappa_3 / T_m ) + \kappa_{2}^{'}]}
\end{equation}
where $\rho$ is the density of liquid water, $R_{\nu}$ is the specific gas constant for water vapor, and  $\kappa_2^{'}$, and $\kappa_3$ are physical constants that characterize the atmosphere. For this analysis we use $\kappa_3=3.739\pm0.012~10^5~\text{K}^2~\text{mb}^{-1}$, and $\kappa_{2}^{'}=22.1~\pm2.2\text{~K~mb}^{-1}$ \citep{Bevis94}. $T_{m}$ is the weighted mean temperature of the atmosphere and is defined as \citep{Davis85}

\begin{equation}
  T_m = \frac{\int ( P_\nu / T)dz}{ \int (P_\nu \ T^2)dz }
\end{equation}
where $P_\nu$ is the partial pressure of water vapor. \cite{Bevis92} present an approximation to $T_m$ based on the surface temperature given by
\begin{equation}
  T_m \approx 70.2 + 0.72~T_s. \label{eq:bevistm}
\end{equation}
where $T_s$ is the surface temperature in kelvin.  This relationship was derived from 8718 radiosonde profiles spanning at two-year period from sites across the United States.  The latitude range of this sample was $27^{\circ}$ to $65^{\circ}$ and a height range of 0 to 1.6~km (Rubin Observatory sits at an altitude of roughly 2.6~km).  \citet{Davis85} find that this relationship has a relative error of less than 2$\%$. Indeed, \citet{Wang16} find that even globally, the approximation of Eq.~\ref{eq:bevistm} produces a relative error on derived PWV of only $1\%-3\%$ with $92\%$ of test sites having a relative error smaller than $2\%$. A flow chart illustrating the steps to go from GNSS observation files to PWV is presented in Fig. \ref{fig:flowchart}.

\subsection{PWV Measurement Uncertainty}\label{sec:pwvuncertainty}

  PWV is computed as $\Pi \cdot (\text{ZTD} - \text{ZHD})$, so the error is composed of three parts; the error on $\Pi$ ($\sigma_{\Pi}$), the error on ZTD ($\sigma_{\text{ZTD}}$), and the error on ZHD ($\sigma_{\text{ZHD}}$).  \citet{Bevis94} demonstrate that the relative error in $\Pi$ is closely approximated by the relative error in $T_m$. So we can write the uncertainty in PWV ($\sigma_{\text{PWV}}$) as a function of the error on $T_m$ ($\sigma_{T_m}$) assuming the errors are independent as
  \begin{equation}
    \sigma_{\text{PWV}}^{2}
  = \Pi^{2}\left(\sigma_{\text{ZTD}}^{2} + \sigma_{\text{ZHD}}^{2}\right)
  + \text{PWV}^{2}\left(\frac{\sigma_{T_m}}{T_m}\right)^{2}.
  \end{equation}
  We can see that at small PWV, the last term vanishes and $\sigma_{\text{PWV}}$ tends to a floor of $\Pi \cdot \sqrt{\sigma^{2}_{\text{ZTD}} + \sigma^{2}_{\text{ZHD}}}$. In practice we report the random components ($\sigma_{\text{ZTD}}$) as the precision and the systematic parts ($\sigma_{\text{ZHD}}$) as the accuracy.

  The ZHD error budget depends on the accuracy of the pressure sensor, the uncertainty in the height difference between the GNSS antenna and the weather station, and uncertainty in the Saastamoinen model itself. Assuming a pressure uncertainty of 0.25~hPa and a height difference uncertainty of 1~m (0.09~hPa), the sensor and height terms contribute 0.57~mm and 0.21~mm
to $\sigma_{\text{ZHD}}$, respectively. The Saastamoinen model is by far the
dominant term: \citet{Feng20} find an RMS of 1.7~mm with respect to
radiosonde ray-tracing. Combining the three terms in quadrature gives
$\sigma_{\text{ZHD}} \approx 1.8$~mm, which corresponds to
$\sigma_{\text{PWV}} \approx 0.27$~mm (0.26~mm from the model term alone). The character of ZHD error is essentially a constant offset or slowly varying and is treated as a systematic contribution to the accuracy (Section \ref{sec:IGScomparison}).

The ZTD uncertainty depends on satellite orbit and clock products, higher-order ionosphere effects, receiver
and antenna hardware noise and stability, multipath error, the mapping function and higher order troposphere
variations, and the process noise set in the Kalman filter. 
The choice of orbit and clock products and of fixed versus float ambiguities has been found to have little impact on PPP ZTD \citep{Stepniak22}.
The formal uncertainty on ZTD reported by gLAB is derived from the covariance
matrix of the extended Kalman filter \citep{gLAB} and reflects the
precision under the assumed model. It does not capture unmodeled systematic
effects such as multipath or unmodeled higher order atmospheric terms, which are time-correlated
rather than random. Neglecting time-correlated errors is known to make
covariance estimates overly optimistic \citep{ElRabbany03}, and some software
scales the reported uncertainties to account for this \citep{Banville21}.
gLAB does not apply such a scaling.
However, the standard deviation of the difference between the gLAB and IGS
ZTD (Section \ref{sec:IGScomparison}) is comparable to the median of the
formal uncertainties reported by gLAB and by IGS combined in quadrature, so we
quote the gLAB formal uncertainty as reported as our estimate of precision.

\section{Data}\label{sec:data}

\subsection{IGS}\label{sec:IGS}

We performed an extensive analysis of gLAB's estimate of position coordinates and the ZTD using data provided by the International GNSS Service (IGS) and Crustal Dynamics Data Information System (CDDIS)\footnote{\url{https://www.earthdata.nasa.gov/centers/cddis-daac}}. IGS is a collection of agencies, universities, and research institutions that work together to provide quality GNSS data products \citep{IGS}. IGS provides free and open access to many products including GNSS satellite ephemerides, Earth rotation parameters, and importantly, ZTD estimates\footnote{\url{https://igs.org/products/\#about}}.  IGS and associate analysis centers generate these products from ground-based GNSS data archived in CDDIS. CDDIS maintains an archive that includes IGS derived products but also observation files, navigation logs, and other raw GNSS data from a network of over 500 ground-based receivers \citep{CDDIS}.

The quality of precise clock and orbit files directly impacts PPP results. The IGS Final products are the most consistent of all IGS products and have the highest quality.   The Multi-GNSS Pilot Project (MGEX) was started in 2016 to analyze all available GNSS signals \citep{IGS-MGEX, MGEX}. Analysis centers work to develop software capable of estimating inter-system calibration biases in addition to tracking, collocating and analyzing multi-constellation signals.  MGEX products by different analysis centers vary in terms of accuracy and consistency and are an active subject of research \citep{Bahadur18,Angrisano20, Bu21, Hou23}. We use only the IGS Final orbit and clock files and examine solutions using only the GPS constellation.

IGS also provides troposphere products\footnote{\url{https://igs.org/products/\#troposphere}}.
These include the ZTD and north and east troposphere gradient components.  Final troposphere estimates are generated by the US Naval Observatory (USNO) analysis center \citep{Byram11} for selected sites in the IGS network and are generally available three weeks after observation date\footnote{\url{https://www.earthdata.nasa.gov/data/space-geodesy-techniques/gnss/atmospheric-products}}.
This is because the software uses the IGS Final precise files.  The troposphere products are computed using a PPP method with the Bernese GNSS software \citep{bernese15}. They use a 27-hour analysis window, a cutoff elevation angle of $7^\circ$ and a temporal resolution of 5 minutes\footnote{\url{https://files.igs.org/pub/resource/pubs/workshop/2012/Poland\%202012\%20-\%20P06\%20Byram\%20PO33.pdf}}.
They also use the Global Mapping Function \citep{GMF06} for the troposphere mapping.  These troposphere products will be used to evaluate gLAB's performance.

\begin{figure}
  \centering
  \includegraphics[width=0.8\textwidth]{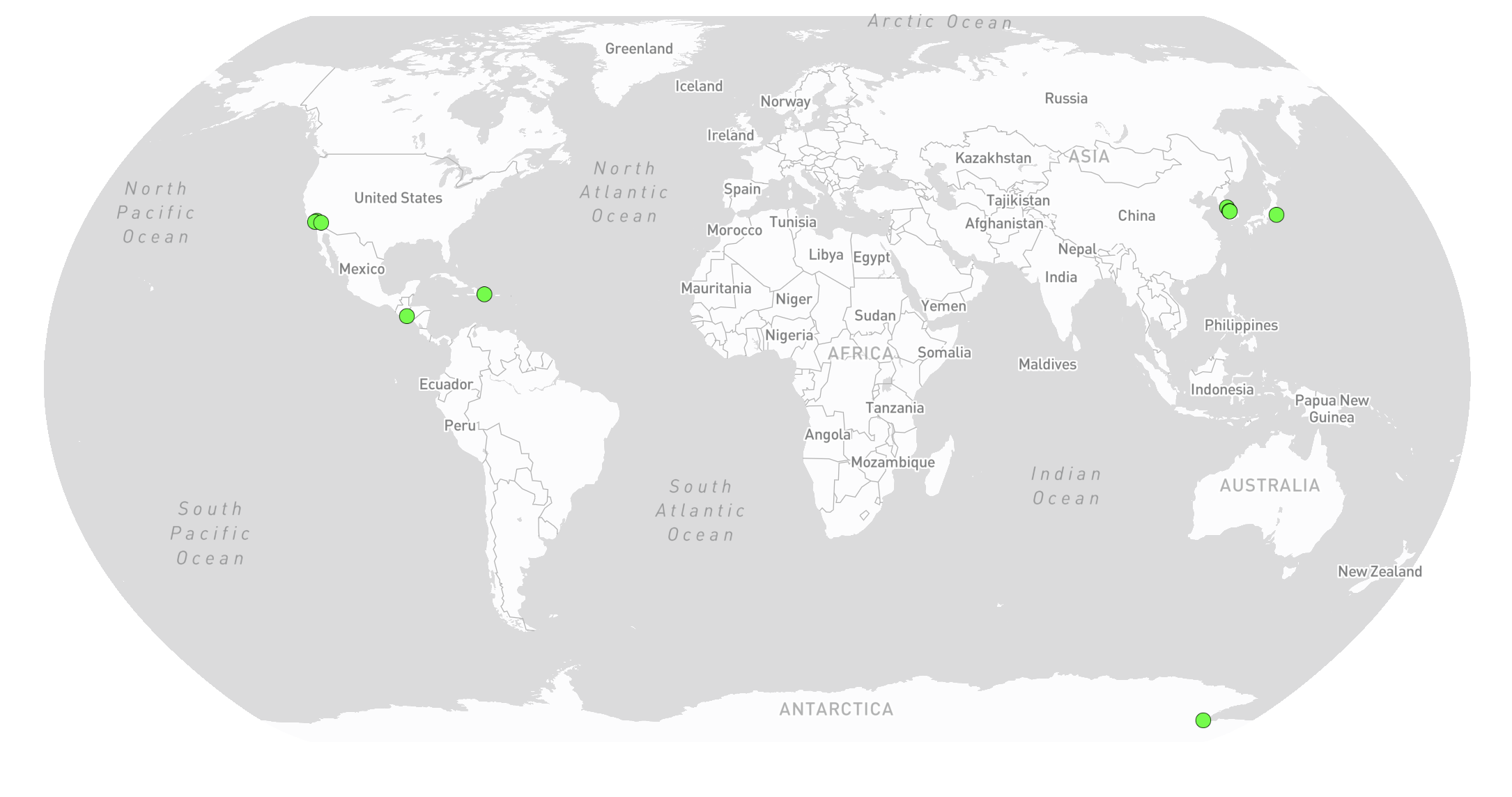}
  \caption{IGS Stations that utilize the Trimble Net R9 receiver.  Station and antenna information is recorded in Table~\ref{table:IGSstations}.  Nine stations are located in the Northern Hemisphere with several stations clustered in South Korea and in the USA. COTE is the only station in the Southern Hemisphere and is located in Antarctica where the poles exhibit unique geometric challenges. Source: \url{https://network.igs.org}}
  \label{fig:stationlocations}
\end{figure}

We compared the (x, y, z) coordinates and ZTD from running gLAB in static PPP mode for 10 IGS stations to the reference values published on the CDDIS web server\footnote{\url{https://cddis.nasa.gov/archive/gnss/data/daily/2025/}}.
These stations were chosen because they record multi-constellation multi-frequency GNSS data with the same receiver we have deployed at Rubin, although only three of these stations also use the same antenna we are using (see Section~\ref{section:AO}). Station information is summarized in Table~\ref{table:IGSstations} and are plotted in Figure~\ref{fig:stationlocations}.  We choose a processing strategy similar to USNO for comparison. However, we decided to employ an analysis window of 3 days and use only the measurements from the center day.  This was to allow enough time for gLAB to converge and to minimize boundary effects.  For each station, four weeks of data were analyzed (2025 May 4--31 May) which included observation files, navigation files, and precise clock products sampled at 30~s intervals.

\begin{center}
\begin{deluxetable}{lllrrrl}
\tabletypesize{\footnotesize}
\tablewidth{0pt}
\tablecolumns{7}
\tablecaption{Summary of IGS Stations That Use Similar Hardware}
\label{table:IGSstations}
\tablehead{\colhead{Station Name} &
        \colhead{Country$/$Region} &
        \colhead{Antenna} &
        \colhead{Latitude} &
        \colhead{Longitude} &
        \colhead{Height (m)} &
        \colhead{Recorded Satellite System}}
\startdata
CIT100USA & USA & TRM57971.00 & 34.1367 & -118.1273 & 215.4  & GPS+GLO+SBAS \\
COTE00ATA & Antarctica & TRM59800.00 & -77.8059   & 161.9978     & 1878.1  & GPS+GLO \\
DAEJ00KOR & South Korea & TRM59800.00 & 36.3994   & 127.3745    & 117.0 & GPS+GLO+GAL+BDS+QZSS\\
MTKA00JPN & Japan & ASH701073.1 & 35.6795 & 139.5614  & 109.0   & GPS+GLO \\
RDSD00DOM & Dominican Rep. & TRM59800.00 & 18.4614 & -69.9113    & -9.2 & GPS+GLO+GAL+BDS \\
SEJN00KOR & South Korea & TRM59800.00 & 36.5222   & 127.3032   & 181.2   & GPS+GLO+GAL+BDS+QZSS+SBAS \\
SSIA00SLV & El Salvador & TRM59800.80 & 13.6973   & -89.1162    & 664.4   & GPS+GLO+GAL \\
SUWN00KOR & South Korea & TRM59800.80 & 37.2755   & 127.0542    & 83.9    & GPS+GLO+GAL+BDS+QZSS+SBAS \\
TABL00USA & USA & TRM57971.00 & 34.3819  & -117.6783  & 2228.0 & GPS+GLO+SBAS \\
WIDC00USA & USA & TRM57971.00 & 33.9348 & -116.3918  & 445.0  & GPS+GLO+SBAS \\
\enddata
\end{deluxetable}
\end{center}

\subsection{Allegheny Observatory}\label{section:AO}
The equipment chosen to obtain GNSS derived PWV measurements at Rubin are a Trimble NetR9 GNSS Reference Receiver and Trimble Zephyr 2 Geodetic antenna. This multi-frequency receiver tracks all current and near-future GNSS signals including GPS L1/L2/L5, GLONASS L1/L2/L3, Galileo E1/E2/E5/E6, and BeiDou B1/B2/B3.  It has sub-millimeter phase-center repeatability, low-elevation tracking, and reduced ground-based multipath.

In order to test the GNSS receiver and antenna, as well as obtain a dataset to develop the necessary analysis pipelines, we commissioned the system at the University of Pittsburgh's Allegheny Observatory (AO) located in Riverview Park in Pittsburgh, Pennsylvania. The observatory sits on a hill 370~m above sea level and avoids surrounding terrain obstruction which could impair precision.
However, the observatory has three domes which comprised the main physical obstructions that could affect the received signal.
The antenna was mounted on a temporary base on the roof of the observatory roughly equidistant between the two domes located on the southern edge of the observatory.
The temporary base was not attached to the building and was instead weighed down with concrete blocks.
We found no evidence that the base moved during the month it was on the roof.
Data from the antenna was transmitted to the receiver, housed inside the observatory, via coaxial connection.

We recorded approximately 1 month of data from 2025 March 10 to April 9.
We set the elevation mask to $10^{\circ}$.
At this time we are recording signals from all available constellations although we are only using GPS in this analysis because of the better availability of precise clock and orbit files.

\subsection{Rubin Observatory}
\label{sec:Rubin_installation}

\begin{figure}
\plottwo{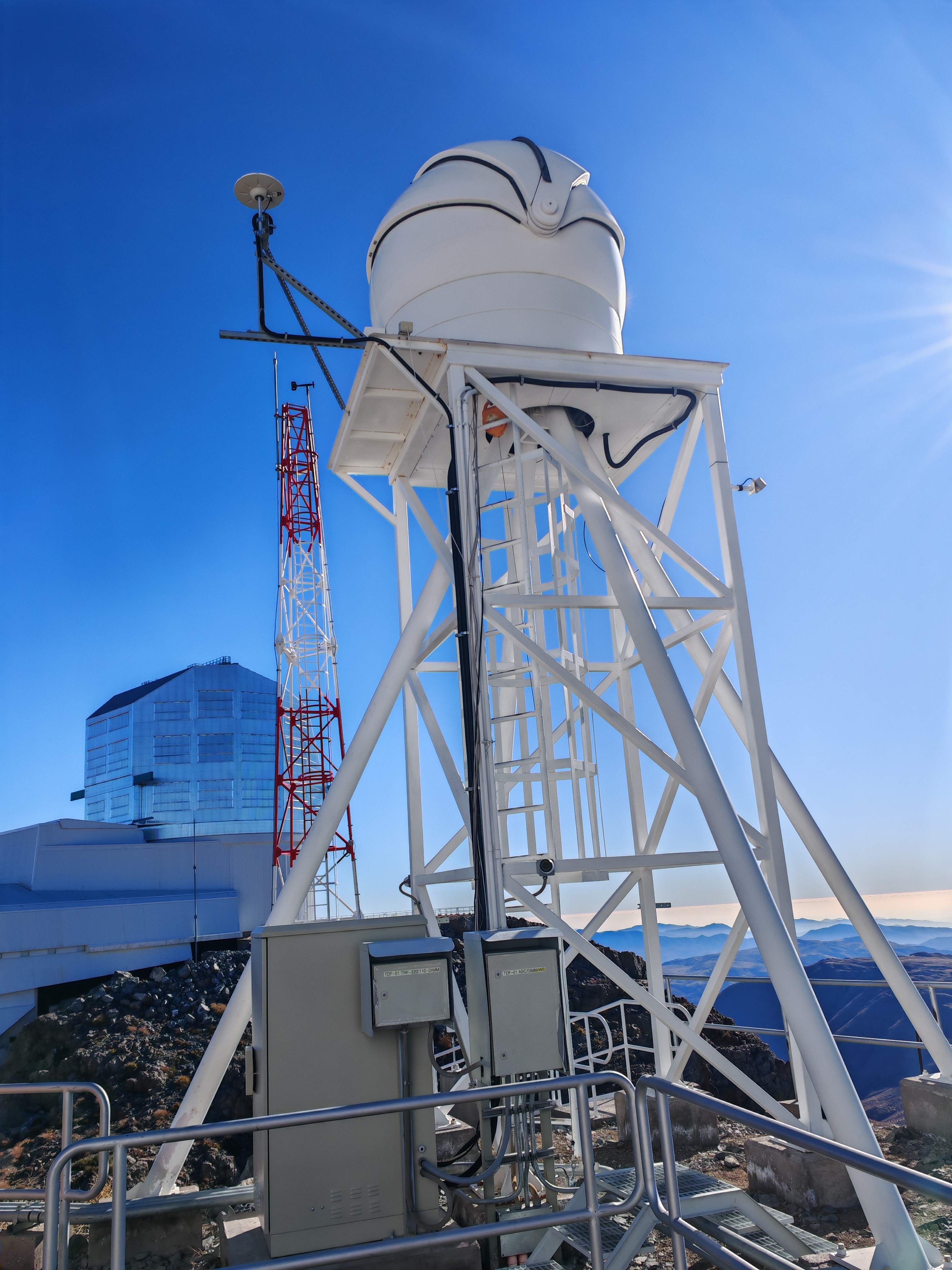}{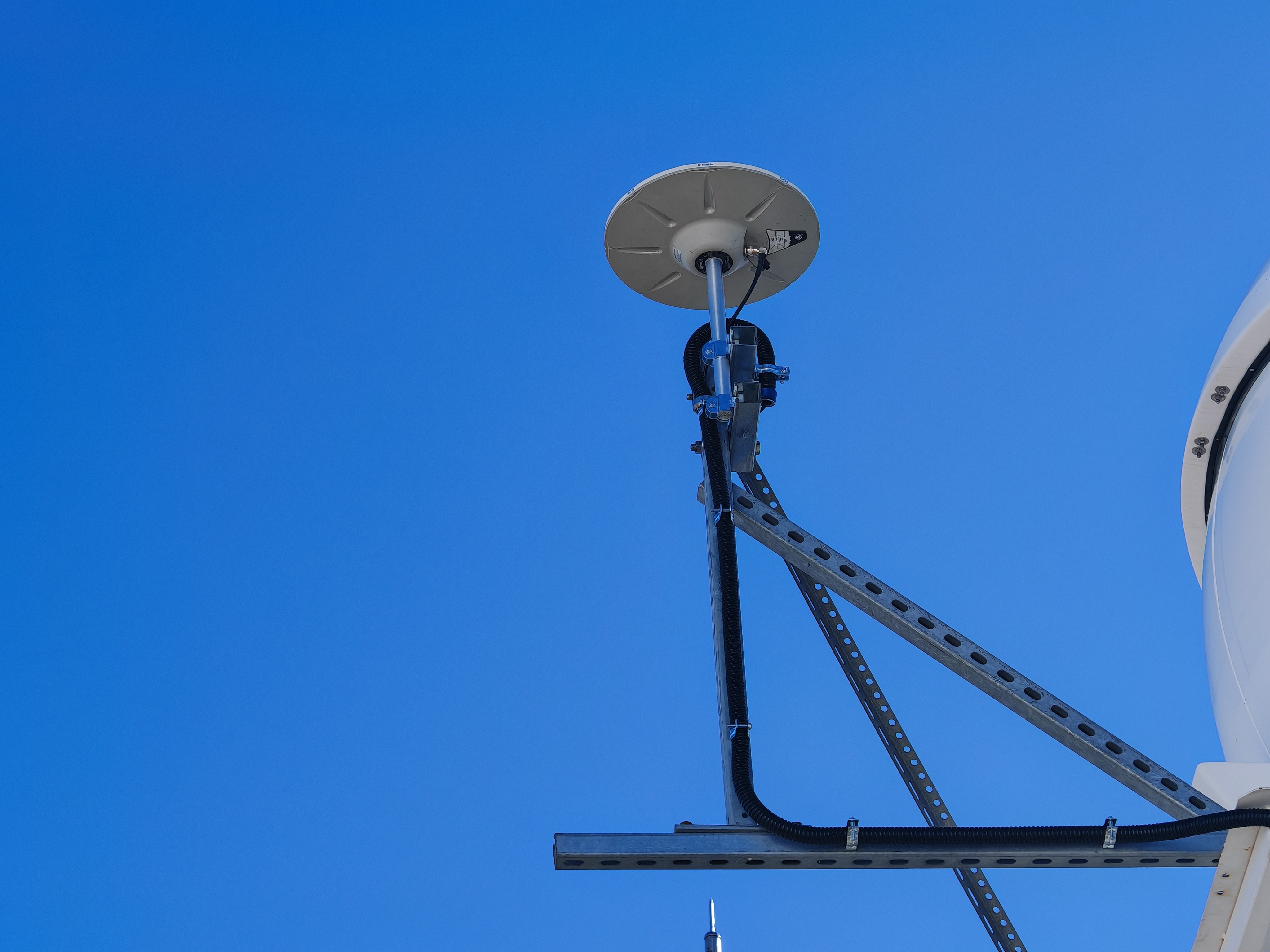}
  \caption{The Zephyr Geodetic 2 antenna was mounted to the tower of the Differential Image Motion Monitor (DIMM) (left)
on a threaded aluminum pipe mounted to an angle bracket made of galvanized steel (right).
A 50~Ohm coax cable was run from the receiver down through a cable wrap to the equipment box near the ground.
The NetR9 receiver is housed in the electrical box in the bottom left and power-over-ethernet for both its power and network connections.
}
  \label{fig:GNSS_installation}
\end{figure}

\begin{figure}
\plottwo{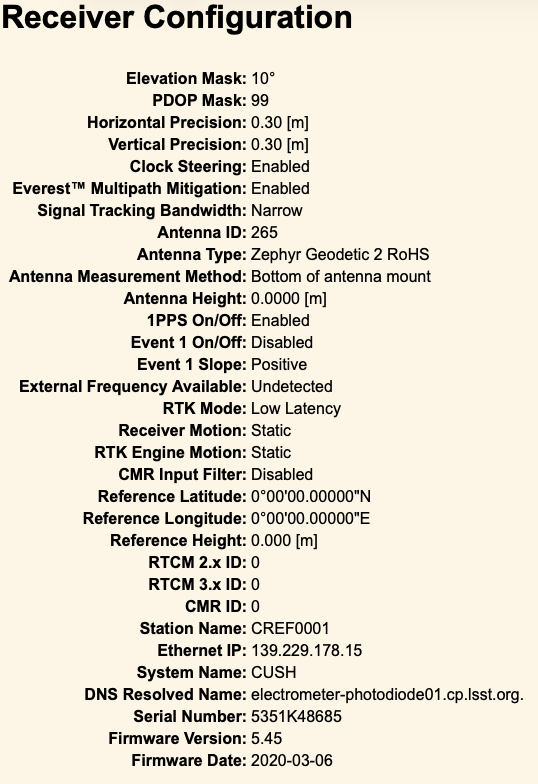}{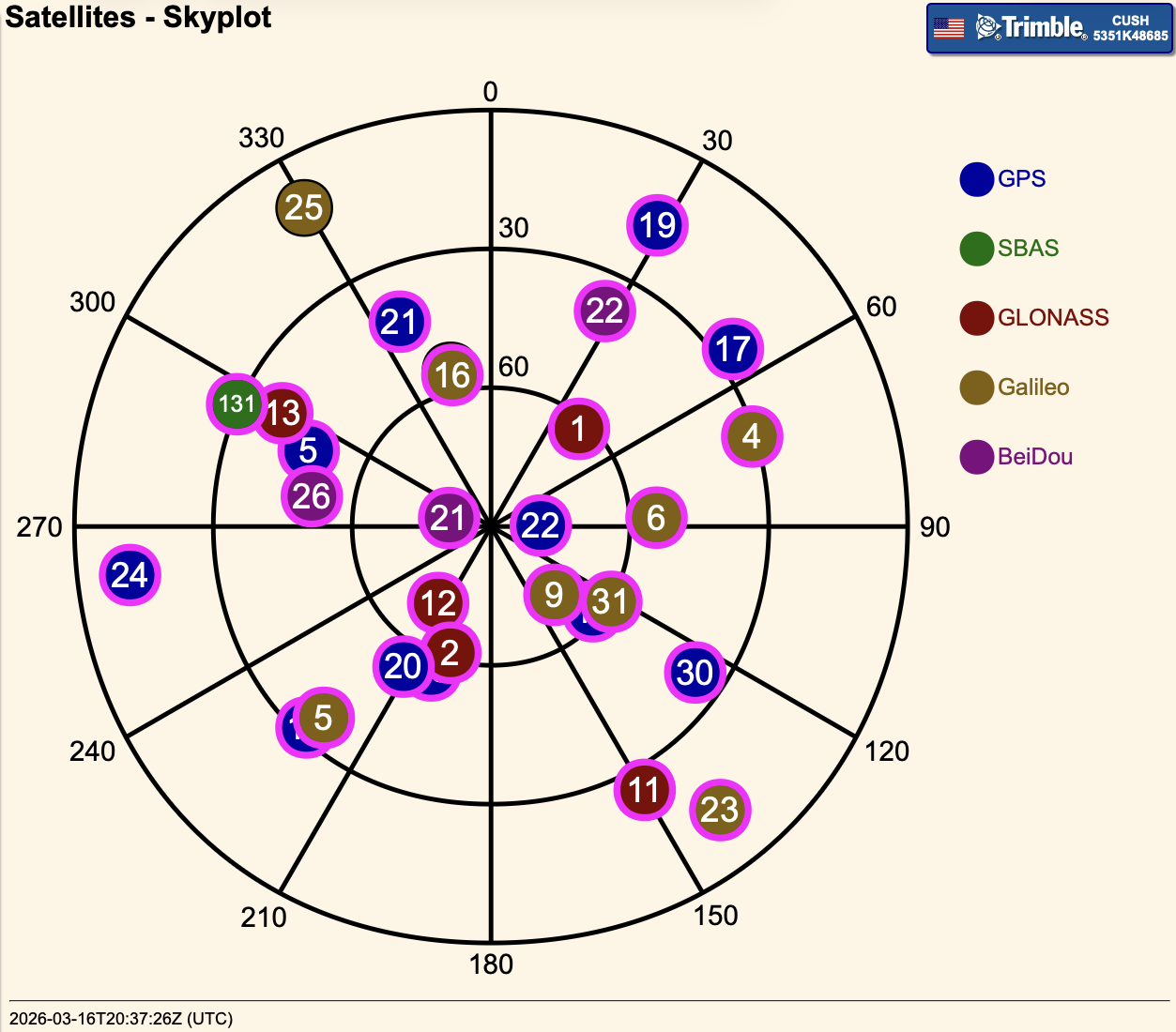}
   \caption{Trimble NetR9 receiver configuration information from the web interface to the device.
   The NetR9 receiver is running firmware 5.45.
   (left) Configuration used to record data from all the constellations.
   (right) Example sky plot as generated by Trimble receiver software showing satellites from all five constellations (GPS, SBAS, GLONASS, Galileo, BeiDou) being visible at a particular moment.
   The receiver records all of the available constellations, although in this present paper we are focused on the GPS signals because those have the best refined timing and orbit information.
   }
   \label{fig:receiverconfiguration}
\end{figure}

Once the AO commissioning phase was successful, we shipped the Trimble GNSS receiver and antenna system down to Chile and it was installed at Rubin Observatory on 2025-09-04 by members of the Rubin Construction Team (Figure~\ref{fig:GNSS_installation}).
The antenna was installed on the tower that supports the Differential Image Motion Monitor (DIMM) system at the summit.
The receiver is in an outside electrical cabinet at the base of the tower and is connected to the antenna via a coax cable and connected to power and the internet via a power-over-ethernet (PoE) connection.
Since installation, the system has been running unattended with no maintenance or interventions necessary.
The system is configured through a web interface hosted by the receiver.
We show the receiver configuration page in Figure~\ref{fig:receiverconfiguration}.
Data are downloaded as nightly files (for a 24-hour UTC day) through the web interface and saved as RINEX v2.11 files.

\subsection{ERA5}

As a sanity check on our measurements, we utilize the European Centre for Medium-Range Weather Forecasts (ECMWF) fifth-generation reanalysis (ERA5) \citep{Hersbach20era5}. This comprehensive global climate dataset includes a large quantity of atmospheric, land, and oceanic parameters from 1940 to the present day.  GNSS measurements are not used to estimate water vapor in their modeling so the ERA5 dataset provides an independent check of our methods.

\subsubsection{ERA5 and Allegheny Observatory}
We use the total column water vapor (TCWV) from the ERA5 hourly data \citep{Hersbach23single} to compare to our GNSS-based PWV measurements from AO.
We also utilize the surface pressure and temperature measurements as we did not yet have a weather station at AO.
\citet{Huang23} evaluate the accuracy of GNSS PWV retrieval with ERA5 in China against radiosonde data and find a bias of -0.43~\PWVunits{} of PWV.
We acquired the data using the Copernicus Climate Change Service (C3S) Climate Data Store (CDS) \citep{copernicussingle}.

Importantly, the height system of ERA5 is based on geopotential while GNSS uses geodetic/ellipsoidal height and weather stations typically record data using the orthometric height (true elevation).  To compare data across these systems we have to reconcile the different height systems \citep{Wang16, Huang23}. At AO, we converted the ERA5 geopotential to orthometric height.  We then used the Geoid Height Calculator provided by NSF GAGE Facility\footnote{\url{https://www.unavco.org/software/geodetic-utilities/geoid-height-calculator/geoid-height-calculator.html}} to convert to ellipsoidal height using the Earth Gravitational Model 2008 (EGM08) \citep{Pavlis12}.
The ERA5 geoid height and geopotential height at the latitude and longitude of AO are 34.61~m and 336.608~m, respectively.
We find that the GPS receiver is $\Delta H=46.47$~m above the ERA5 surface.

The ZHD model is sensitive to the pressure.
We therefore adjust the pressure and temperature from ERA5 to the height of the receiver using the barometric formula and the standard temperature lapse rate $L$ of $-6.5$~K~km$^{-1}$ \citep[see, e.g.,][]{Ssenyunzi20}:
\begin{equation}
  P_{\text{GPS}} = P_{\text{ERA5}}e^{-\frac{M_dg\Delta H}{RT}}
\end{equation}
\begin{equation}
  T_{\text{GPS}} = T_{\text{ERA5}} - L \Delta H
\end{equation}
where $T$ is the standard temperature of the atmosphere at sea level, $M_d$ is the molar mass of dry air, $g$ is 9.80665~m~s$^{-2}$, $R$ is the universal gas constant for air, $T_{\text{ERA5}}$ and $P_{\text{ERA5}}$ are the surface temperature and pressure from ERA5 and $T_{\text{GPS}}$ and $P_{\text{GPS}}$ are the air temperature and pressure at the GPS site. Finally, we adjust the TCWV to the GPS receiver height following \citet{Kouba08}.

\subsubsection{ERA5 and Rubin Observatory}
We also compare the ERA5 data to Rubin Observatory.
This dataset covers the entire Earth on a 31~km model grid at 37 pressure levels (which correspond to heights) and provides, among other things, estimates of temperature and specific humidity at each pressure level and corresponding grid points.  ERA5 grid points are a consequence of the model's discretization and do not represent more precise or privileged locations; they represent the native spatial resolution of ~31 km, which determines the scale over which atmospheric conditions are resolved. This dataset was also acquired using C3S CDS \citep{copernicus}\footnote{Both ERA5 datasets are released for use under the CC-BY license.}.

We calculate PWV following a method outlined in \cite{Bock19} and \cite{Ssenyunzi20} and integrate the specific humidity, $q$, over pressure from the surface at Rubin Observatory $P_s$ to the top of the atmosphere (0~hPa):
\begin{equation}
  \text{PWV} = \frac{1}{\rho_wg}\int_{P_s}^{0} q(P)dP
\end{equation}
where $\rho_w$ is the density of water.  We generate the vertical profile of the specific humidity $q(P)$ by bilinearly interpolating between the nearest four grid points at each pressure level.  Rubin Observatory lies between two pressure levels so we linearly interpolated between the adjacent pressure levels as well and evaluated the specific humidity according to local measurements of the pressure.  Since our weather station is co-located with the receiver at Rubin Observatory, we use this as our surface pressure measurement rather than using temperature and geopotential estimates from ERA5\footnote{Actually, the weather station is mounted approximately 7~meters below the GNSS antenna.  ZHD is quite sensitive to pressure at the height of the antenna.  We estimate the overall derivative to be $\frac{d \text{PWV}}{\Delta \text{H}}=-0.33~(kg/m^2) / 10~(m)$.}.  This method of estimating PWV is more accurate than the TCWV variable, especially in mountainous regions.

\section{Analysis}\label{sec:analysis}

\subsection{Evaluation of gLAB Performance against IGS}\label{sec:IGScomparison}

\begin{figure}
  \centering
  \includegraphics[width=0.8\textwidth]{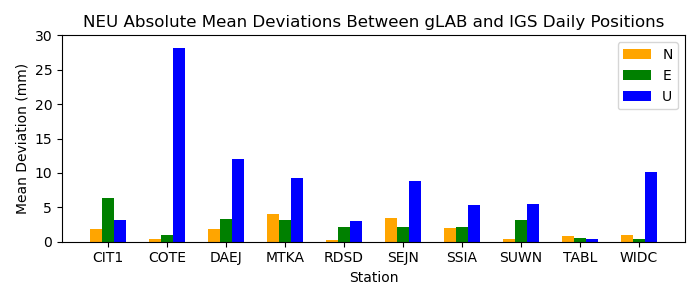}
  \caption{Absolute mean deviations between the gLAB determined location and the IGS daily SINEX position using all epochs with a navigation solution. Results are presented in the north, east, and up directions (NEU). The mean difference for the entire sample between gLAB and IGS NEU position is 1.5~mm (N), -1.7~mm (E), and 5.7~mm (U) when we exclude COTE.  The largest deviations are in the Up direction because satellites are not visible below the horizon. IGS daily positions have an accuracy of 2~mm in horizontal directions and 5~mm in the vertical direction.}
  \label{fig:NEUmean}
\end{figure}

We compared receiver coordinates determined by gLAB to the station coordinates in the daily IGS SINEX solutions, a standard text format in which the IGS distributes its estimated positions and their covariances, in the WGS84 system. It is customary to evaluate position residuals in terms of the North, East, and Up components (NEU).  The coordinates estimated at 5-minute intervals for each day were compared to the daily IGS position for each station.  The mean deviation, standard deviation, and Root Mean Square Error (RMSE) were calculated for each station using every available epoch and are listed in Table~\ref{table:NEUmean}. The absolute mean deviations are presented in Figure \ref{fig:NEUmean}.

The results of this comparison are that gLAB is recovering IGS positions at the 1--3~mm level. We find the mean difference between gLAB and IGS NEU position to be 1.4~mm, -1.4~mm, and 2.8~mm (see Table~\ref{table:NEUmean}).  IGS daily positions have an accuracy of 2~mm in the horizontal plane and 5~mm in the vertical direction\footnote{\url{https://igs.org/products/}\label{igs}}.  The larger deviations in the Up direction are expected due to the geometry of constraining position in a plane from satellites from all directions versus constraining position along a line where we can only have satellites above the horizon and not below.

\begin{center}
\begin{deluxetable}{lrrrrrrrrr}
\tabletypesize{\footnotesize}
\tablewidth{0pt}
\tablecolumns{10}
\tablecaption{NEU Position Statistics in mm}
\label{table:NEUmean}
\tablehead{ \colhead{} & \multicolumn{3}{c}{North} & \multicolumn{3}{c}{East} & \multicolumn{3}{c}{Up}\\
  \colhead{Station Name} &  \colhead{Mean} &  \colhead{$\sigma$} &  \colhead{RMSE} &\colhead{Mean} &  \colhead{$\sigma$} &
  \colhead{RMSE} &\colhead{Mean} &  \colhead{$\sigma$} &  \colhead{RMSE}}
\startdata
CIT1 & 1.8 & 2.7 & 3.2 & -6.3 & 3.6 & 7.3 & -3.1 & 5.7 & 6.5 \\
COTE & 0.3 & 3.7 & 3.7 & 1.0 & 6.3 & 6.4 & -28.2 & 7.1 & 29.1 \\
DAEJ & 1.8 & 1.5 & 2.3 & -3.3 & 1.6 & 3.7 & 12.1 & 6.8 & 13.9 \\
MTKA & 4.0 & 3.1 & 5.1 & -3.1 & 4.4 & 5.4 & 9.2 & 7.7 & 12.0 \\
RDSD & 0.3 & 4.6 & 4.6 & 2.2 & 5.6 & 6.1 & 3.0 & 9.8 & 10.2 \\
SEJN & 3.5 & 1.8 & 3.9 & -2.2 & 2.2 & 3.1 & 8.8 & 5.2 & 10.2 \\
SSIA & 2.0 & 3.8 & 4.3 & 2.2 & 5.3 & 5.7 & 5.4 & 9.4 & 10.8 \\
SUWN & 0.4 & 1.8 & 1.8 & -3.2 & 1.9 & 3.8 & 5.5 & 6.0 & 8.1 \\
TABL & 0.8 & 1.6 & 1.7 & -0.6 & 2.7 & 2.8 & 0.4 & 5.3 & 5.3 \\
WIDC & -0.9 & 2.9 & 3.0 & -0.4 & 3.8 & 3.8 & 10.2 & 7.4 & 12.6 \\
\tableline
All & 1.4 & 3.3 & 3.5 & -1.4 & 4.8 & 5.0 & 2.8 & 12.8 & 13.0 \\
No COTE & 1.5 & 3.2 & 3.5 & -1.7 & 4.5 & 4.8 & 5.7 & 8.6 & 10.3 \\
\enddata
\end{deluxetable}
\end{center}

The COTE IGS site is located in Antarctica and experiences the largest deviations in the Up direction.  The natural geometric effect is further exacerbated by the orbits of the GNSS constellations with prioritize mid-latitudes over polar latitudes.  GNSS constellations are designed to maximize the number of satellite visible from any point on Earth in the latitude ranges where people live.  Thus, they do not include polar orbits. For receivers near the poles, GNSS satellites will never reach high elevations, increasing uncertainty in position in the vertical direction.  Additionally, around $20$\% of epochs have no solution because there are not enough satellites in view. We find the mean difference between gLAB and IGS NEU position to be 1.5~mm, -1.7~mm, and 5.7~mm when we exclude COTE.

We find NEU RMSE values of 3.5~mm, 4.8~mm, and 10.3~mm when we exclude COTE.  This is comparable to \cite{Vazquez23} who find values of 6.1~mm, 8.4~mm and 11.7~mm respectively. We may have better results due to our longer convergence time from our larger analysis window.

\begin{figure}
  \centering
  \includegraphics[width=0.8\textwidth]{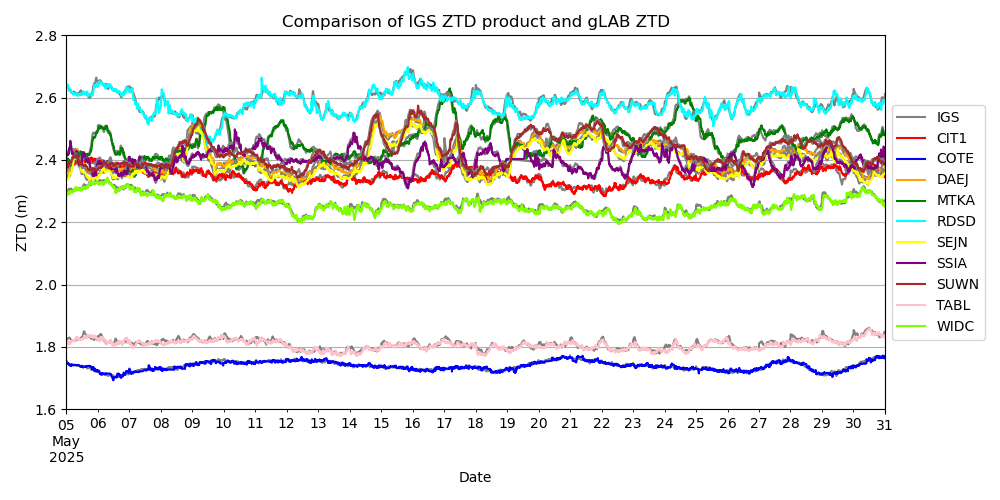}
  \caption{ZTD from IGS at 10 stations used in this comparison (gray) from May 5-31, 2025. Over-plotted is the ZTD calculated from gLAB. Our sample is diverse with dry conditions in Antarctica (COTE) and more variable PWV in the Dominican Republic (RDSD) and South Korea (DAEJ, SUWN, SEJN).}
  \label{fig:ZTDdaily}
\end{figure}

\begin{figure}
  \centering
  \includegraphics[width=0.8\textwidth]{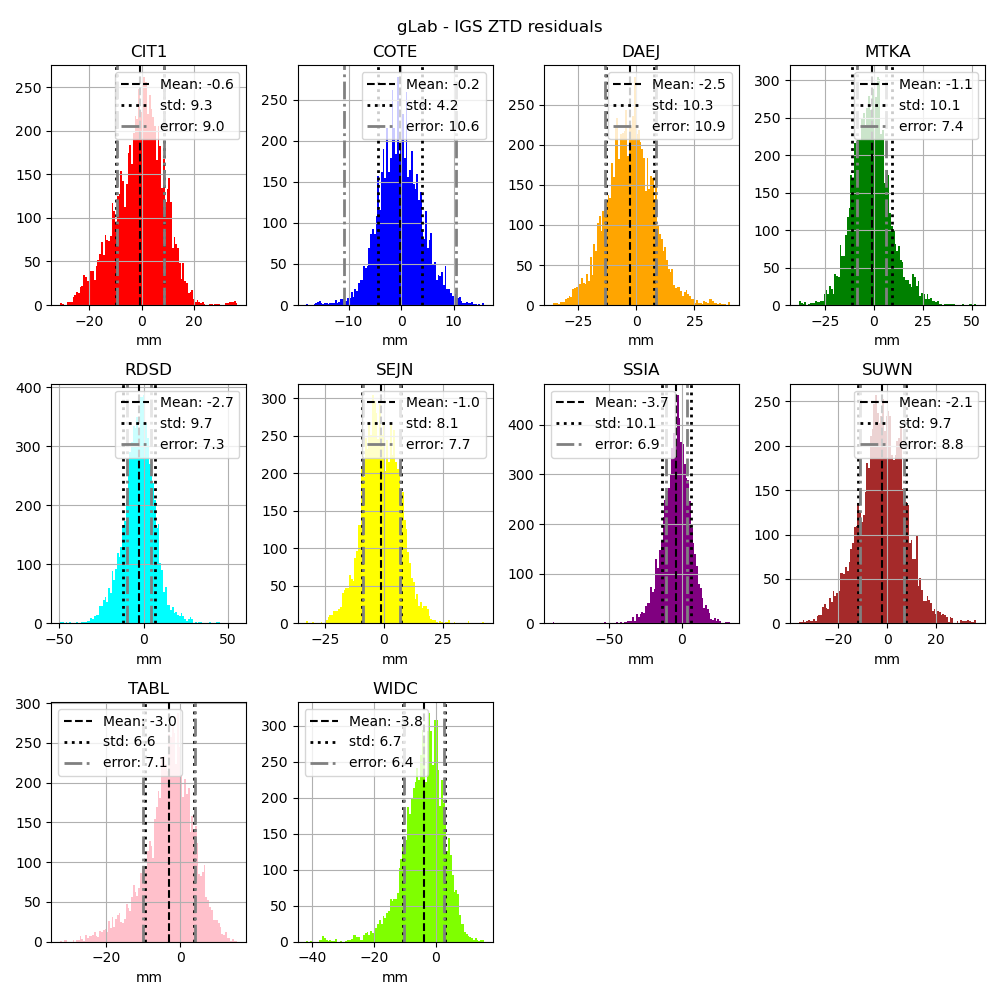}
  \caption{The gLAB-IGS residuals in ZTD for 10 IGS stations.  The mean difference for all stations is 2~mm. For reference, the ZTD accuracy of IGS is 4~mm. The two major differences in processing between gLAB and the IGS ZTD product is the choice of mapping function and that gLAB does not estimate troposphere gradients.  The standard deviation of the distribution of residuals is well represented by the median error calculated from gLAB and IGS with the exception of COTE.  We suspect that this is due to its polar location and that gLAB may overestimate the uncertainties due to satellite geometry.}
  \label{fig:ZTDdiffhistogram}
\end{figure}

We next compared gLAB's measurement of ZTD with the IGS troposphere products.
We first removed ZTD outliers by requiring that the formal error $\sigma_{\text{ZTD}}$ is less than the median $\sigma_{\text{ZTD}} + 3.5 \times$ standard deviation of $\sigma_{\text{ZTD}}$. We then matched epochs between gLAB and IGS. These results are plotted in Figure \ref{fig:ZTDdaily}.  Residual histograms are presented in Figure \ref{fig:ZTDdiffhistogram}.  An estimate of the typical residual error is $\sqrt{(\sigma_{\text{IGS}}^{2} + \sigma_{\text{gLAB}}^{2})}$ where $\sigma_{\text{IGS}}$ and $\sigma_{\text{gLAB}}$ are the median errors of the sample.
The standard devation of the residuals is well represented by the reported uncertainties.  The mean difference between gLAB and IGS for the entire sample is 2~mm in ZTD.

Final tropospheric estimates from IGS have an accuracy of 4~mm.$^{\ref{igs}}$. The hydrostatic delay used to convert ZTD to PWV contributes a
further uncertainty of 1.8~mm (Section~\ref{sec:pwvuncertainty}) Combining the IGS
accuracy and the hydrostatic delay uncertainty in quadrature, and adding the
2~mm bias linearly, gives a conservative estimate of the total accuracy of
6.4~mm.  This translates to approximately $1$~\PWVunits{} in PWV.  The bias could be due to a difference in processing strategies between gLAB and IGS.  Notably, USNO models north and east troposphere gradients and employs a different mapping function.

The RMSE is again used to evaluate gLAB performance and is calculated in this instance as
\begin{equation}
  \text{RMSE} = \sqrt{\frac{\sum_{i=1}^{n} (\text{ZTD}_{\text{gLAB},i} - \text{ZTD}_{\text{IGS},i})^2}{n}}
  \end{equation}
over all epochs $i$. These results are presented in Table~\ref{table:RMSEZTD}. We find a overall RMSE of 9.1~mm, less than 1~cm,  which is an improvement over results in \citet{Mendez18}, who find typical RMSE values of several centimeters. We are using an updated version of gLAB which may account for some of the difference. gLAB performs best (i.e. lowest RMSE) at the three locations with the lowest ZTD values. We find an RMSE at COTE of 4.2~mm, TABL of 7.3~mm, and WIDC with 7.7~mm.  Characteristic values of ZTD at Rubin Observatory are around 1.75~m, similar to COTE.

\begin{center}
\begin{deluxetable}{lr}
\tabletypesize{\footnotesize}
\tablewidth{0pt}
\tablecolumns{3}
\tablecaption{ZTD RMSE}
\label{table:RMSEZTD}
\tablehead{
  \colhead{Station Name} & \colhead{RMSE} \\
& (mm)}
\startdata
CIT1 & 9.3 \\
COTE & 4.2 \\
DAEJ & 10.6 \\
MTKA & 10.1 \\
RDSD & 10.1 \\
SEJN & 8.1 \\
SSIA & 10.8 \\
SUWN & 9.9 \\
TABL & 7.3 \\
WIDC & 7.7 \\
All & 9.1\\
\enddata
\end{deluxetable}
\end{center}

The quality of the observations affects the resulting position and ZTD errors differently. In static PPP mode, the position coordinates are assumed to be constant parameters over the analysis window (72 hours) and are less sensitive to short data gaps and transient errors like multipath. The wet tropospheric delay in gLAB is modeled by a random walk process.  Consequently, ZTD estimation is sensitive to cycle slip detection and ambiguity resolution and to error sources that have sub-daily variability.

\cite{Stepniak22} performed a detailed statistical analysis of ZTD outliers in PPP approaches.
They demonstrate that changing the elevation cutoff angle from 3$^{\circ}$ to 10$^{\circ}$ can have a large impact with mean differences increased to 3~mm due to elevation-dependent error sources like multipath and ambiguity fixing of low-elevation satellites. We found the difference between lowering the elevation cutoff angle to 3$^{\circ}$ and increasing to 10$^{\circ}$ resulted in a change in the mean difference of 0.3~mm, indicating gLAB may be less sensitive elevation-dependent errors. It is also probable that the design of the Trimble hardware used in this comparison is more reliable at lower elevations.
We are considering changing our receiver configuration to record data down to elevations of 3$^{\circ}$ as it is easy to filter out during the analyses stage.

Overall we find gLAB's performance and ease of use to be excellent.

\subsection{Processing strategy for Allegheny and Rubin Observatories}\label{sec:processingstrategy}

We process the data with the gLAB configuration outlined in Table~\ref{table:glabconfiguration} with a few adjustments.  We use all available recorded epochs, which gives us a data sampling rate of once every 15~seconds. We also specify an a priori 3D Cartesian Earth-Centered, Earth-Fixed (ECEF) receiver position of x=841926.54~m, y=-4784854.80~m, and z=4119109.24~m for AO and x=1819089.40~m, y=-5208439.98~m, and z=-3195186.44~m for Rubin Observatory. We adjust the process noise for the troposphere estimation to be 25~mm$^2$~h$^{-1}$ following \citet{Stepniak22}.  We continue to use a total analysis window of 72 hours and trim the first and last 24 hours to minimize boundary effects.

Various algorithms in gLAB are employed to reject outliers in phase and code measurements. These screening techniques are not perfect.  Inevitably outliers will propagate through the software and the processing results (i.e., ZTD) will contain large errors or discontinuities.  For AO we reject outliers as described in Section \ref{sec:IGScomparison}. At Rubin Observatory we filter ZTD following the procedure outlined in \citet{Bock20}. We first use a range check on ZTD of 1.5~m to 2~m that is large enough to account for the natural fluctuations in ZTD. The formal errors calculated by gLAB are sensitive to the quality of ZTD and can also be used to screen for outliers.  The empirical probability distribution function of $\sigma_{\text{ZTD}}$ resembles a $\chi^2$ distribution.  We use the 97\% percentile to trim outliers. This translates to discarding measurements with $\sigma_{\text{ZTD}} > 3.4$~mm. Finally, we apply a range check on the formal errors of median $\sigma_{\text{ZTD}}+3.5 \times$ standard deviation of $\sigma_{\text{ZTD}}$. It is important to first trim outliers and then perform the range check as the standard deviation is not a robust estimator.

\begin{figure}
  \centering
  \includegraphics[width=0.8\textwidth]{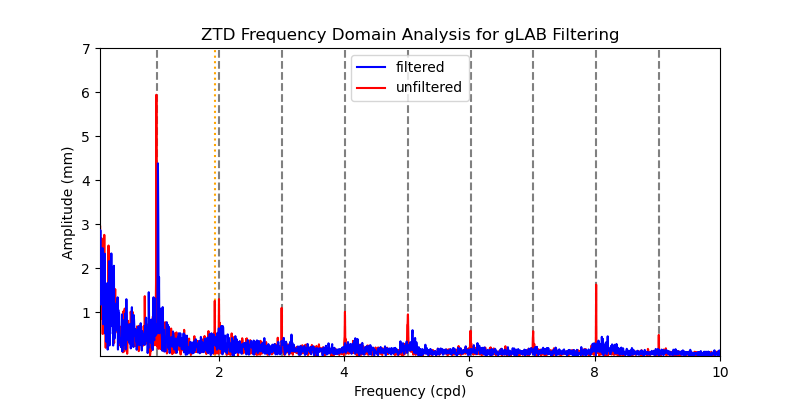}\\
  \caption{Fast fourier transform of ZTD for all of the data (red) and after we screen for ZTD outliers (blue).  The y-axis has units of mm and the x-axis is in cycles-per-day.  The gray dashed lines mark one sidereal day and its harmonics.  The orange dotted line marks the M2 tidal constituent, the primary semi-diurnal lunar component of tides with a period of 12.42~hours. Our screening procedure is effective at removing spurious orbital artifacts appearing in the precise orbit and clock files at harmonics of 1 sidereal day.}
  \label{fig:fft_filtering}
\end{figure}

In Figure \ref{fig:fft_filtering} we plot the ZTD time series amplitude spectra using Fast Fourier Transformation (FFT) for the Rubin data. We linearly interpolated over data gaps. The unfiltered ZTD is plotted in red and the results from the screening process outlined above are plotted in blue.  The gray dashed lines mark harmonics of the K1 tide, the lunisolar diurnal constituent, whose period is one sidereal day.  \citet{Zajdel22} extensively studied orbital artifacts in PPP time series.  They conclude that the existence of the signals at these harmonics is likely in the satellite orbits/clocks and should be recognized as orbital artifacts. We see that our outlier rejection scheme reduces the spurious signals at the K1 frequency and harmonics.  The orange dashed line marks the M2 lunar constituent and is related to the Moon's effect on the tides.  This signal peaks at 12 hours and 25 minutes and is also reduced by our ZTD screening.

\section{Results}\label{sec:results}

\subsection{PWV from Allegheny Observatory Commissioning}

\begin{figure}
  \centering
  \includegraphics[width=1\textwidth]{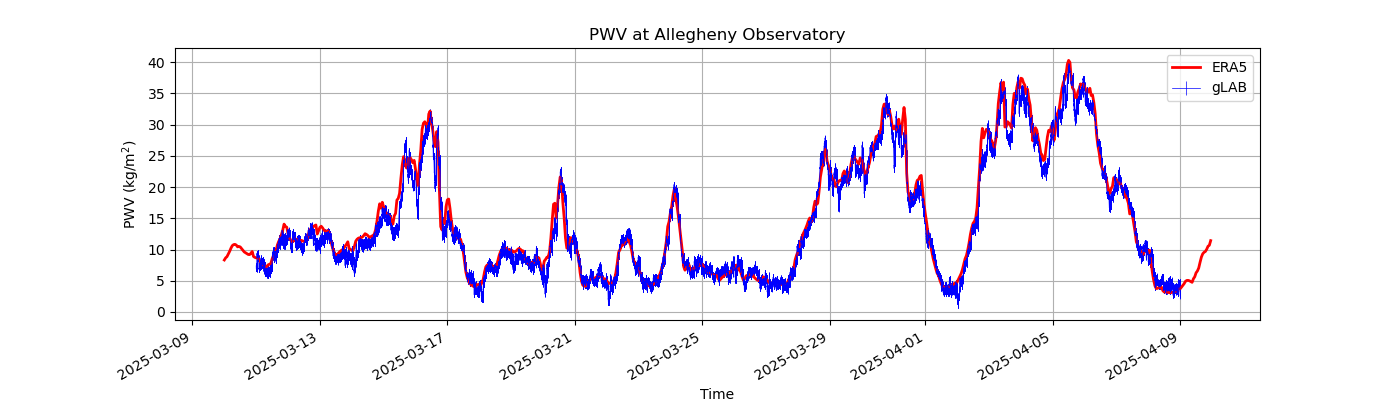}\\
  \caption{PWV at Allegheny Observatory (AO) from March 11, 2025 to April 10, 2025 processed through gLAB (blue) compared with that released by ERA5 (red). PWV at AO is dynamic ranging from less than 5~\PWVunits{} to nearly 40~\PWVunits{}, sometimes varying over 20~\PWVunits{} over the course of a day.  We find the mean difference between AO and ERA5 to be 0.22~\PWVunits{} in PWV.}
  \label{fig:AO_pwv}
\end{figure}

PWV was measured at AO from 2025 March 11 to April 10 and is plotted in Figure~\ref{fig:AO_pwv} in blue. The accuracy of gLAB-PWV was verified by a comparison to ERA5 hourly TCWV, plotted in red. We additionally utilized ERA5 temperature at 2 meters and surface pressure to convert ZTD to PWV; this correlation in weather sources means this comparison is not entirely indpendent.

Allegheny Observatory sits at low altitude in a humid continental climate with four distinct seasons and large seasonal temperature differences.  We therefore expect to see
greater mean PWV values due to the lower altitude and greater atmospheric moisture as compared to Rubin Observatory. We also expect more variability in PWV due to the frequent passage of mid-latitude weather systems.
We find the PWV at AO varies from approximately 5~\PWVunits{} to 40~\PWVunits{}. On April 3rd, PWV increased from about 15~\PWVunits{} to 30~\PWVunits{}, accompanied by 0.84~in of rain over the course of the day\footnote{\url https://www.ncei.noaa.gov/access/past-weather/}.
The time frame from March 18th to March 27th varied the least, only ranging from about 5~\PWVunits{} to 20~\PWVunits{}. This is coupled by decreased rainfall through the region.

\begin{figure}
  \centering
  \includegraphics[width=0.5\textwidth]{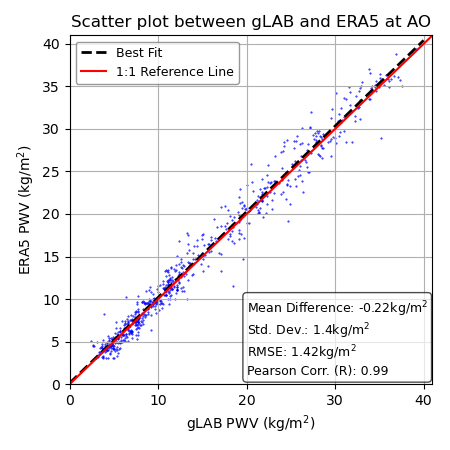}\\
  \caption{A comparison of GNSS-derived PWV and ERA5 PWV at AO.  We find excellent correlation between the two datasets with a Pearson Correlation coefficient of $R=0.99$. An important caveat is that pressure and temperature estimates from ERA5 were used to convert gLAB ZTD to PWV.}
  \label{fig:AO_scatter}
\end{figure}

Figure~\ref{fig:AO_scatter} compares gLAB-PWV and ERA5 TCWV.  We find strong
correlation with a Pearson Correlation coefficient of 0.99. The line of best fit sits just above the 1:1
reference line, with a mean difference between gLAB and ERA5 of -0.22~\PWVunits{}. There is a strong concentration of data points below 10~\PWVunits{}, with less presence as the values increase. The standard deviation is 1.4~\PWVunits{} and the RMSE is 1.42~\PWVunits{}.

\subsection{PWV at Rubin Observatory}

\begin{figure}
  \centering
  \gridline{\fig{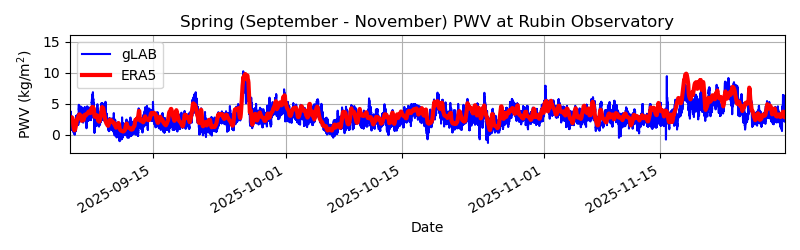}{0.8\textwidth}{\vspace{-15pt}(a)}}
  \vspace{-15pt}
  \gridline{\fig{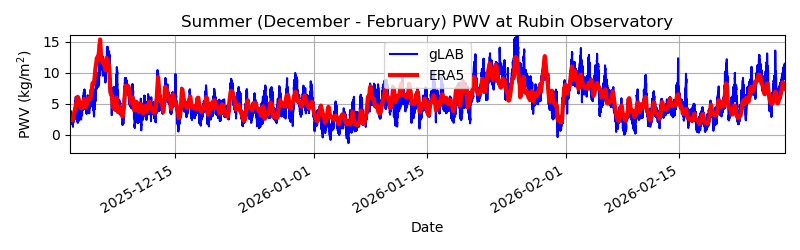}{0.8\textwidth}{\vspace{-15pt}(b)}}
    \vspace{-15pt}
    \gridline{\fig{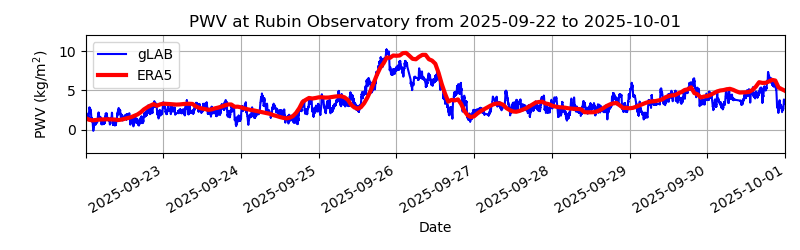}{0.8\textwidth}{\vspace{-15pt}(c)}}
      \vspace{-15pt}
   \gridline{\fig{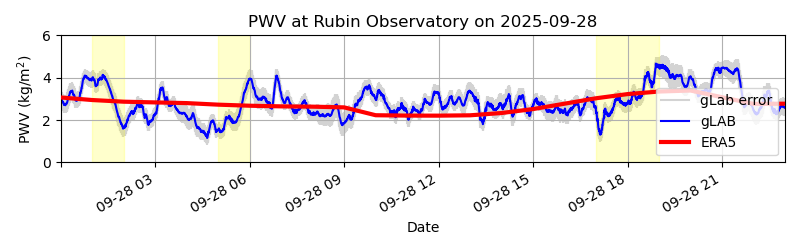}{0.8\textwidth}{\vspace{-15pt}(d)}}
  \caption{Panels (a) and (b): PWV at Rubin Observatory in spring from 2025 September 5 to November 30 and summer from 2025 December 1 2026 January 1. GNSS measurements are in blue and ERA5 in red.  Note that the GNSS measurements are taken much more frequently than the ERA5 model points, so there is more variation at short timescales.
    Panel (c): A zoom in of one week and Panel (d): PWV during a typical spring day. Error bars are plotted in gray. Areas in yellow highlight regions where PWV changes by more than 2~\PWVunits{} in relatively short time period of 1--2 hrs. There is variation in PWV that the ERA5 model cannot resolve.  See Figure~\ref{fig:PSD} for a power-spectrum analysis of the variation in each.
    }
  \label{fig:rubinPWV}
\end{figure}

We present PWV measured at Rubin Observatory from the six months of 2025 September 5 to 2026 February 28 in Figure~\ref{fig:rubinPWV}. PWV is calculated according to the steps outlined in Section~\ref{sec:PWVretrieval} using air temperature and pressure measurements taken at the Rubin Observatory weather station. We found nine gaps in the weather data greater than 5 minutes.  These gaps ranged from 26~minutes to nearly 10 hours. GNSS measurements for these gaps are discarded. Propagating the formal ZTD error yields a typical PWV precision of \PWVprecision{}~\PWVunits{}. ERA5 PWV is overplotted in red. During the spring season (September--November) PWV varies from approximately 0~\PWVunits{} to 10~\PWVunits{} with typical values between 1~\PWVunits{} and 5~\PWVunits{}. The mean value over that period is \springPWVmean{}~\PWVunits{}.  Also presented in Figure~\ref{fig:rubinPWV} are measurements from the summer season, December--February.  We see higher maximum fluctuations of PWV at 15~\PWVunits{} and calculate a mean value of \summerPWVmean{}~\PWVunits{}.  These variations are also captured in the ERA5 weather model.

Sub-daily variations in PWV measured by gLAB are significant and real in comparison to the estimated uncertainty.
We highlight several regions where the variation in PWV is greater than 2~\PWVunits{} over the course of 1--2 hours.

\begin{figure}
  \centering
  \includegraphics[width=0.8\textwidth]{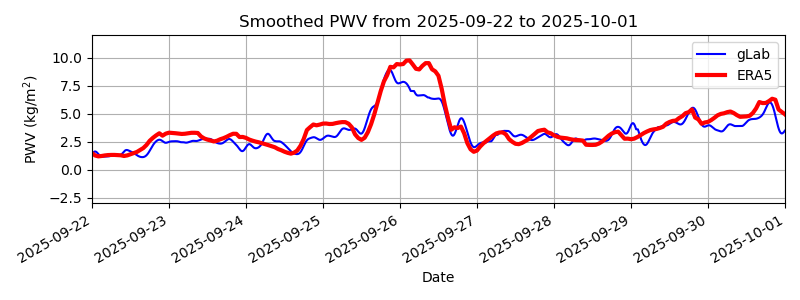}\\
  \caption{GNSS PWV (blue) is smoothed to match the resolution of ERA5 (red). We low-pass filtered the data using a 6-hour Gaussian window. The data are then matched at 1-hour intervals for comparison.}
  \label{fig:rubinPWVsmooth}
\end{figure}

\begin{figure}
  \centering
  \includegraphics[width=0.5\textwidth]{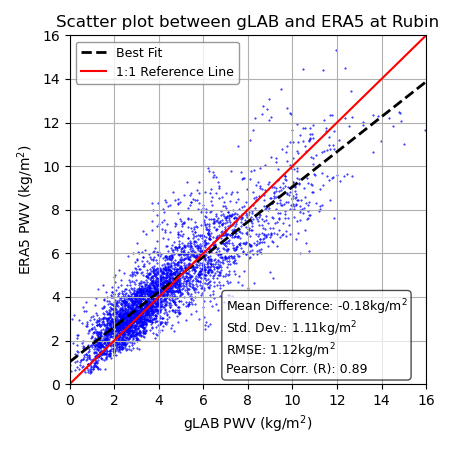}\\
  \caption{ERA5 PWV vs GNSS PWV.  The red solid line marks the 1:1 correlation.  The dashed black line is the best fit line. There is a \biasgLABERA{}~\PWVunits{} bias and a standard deviation of \PWVstdgLABERA~\PWVunits{}. A strong positive correlation exists between the datasets with a Pearson Correlation coefficient of \RgLABERA{} and the RMSE is \PWVRMSEgLABERA~\PWVunits{}.}
  \label{fig:PWVscatter}
\end{figure}

We low-pass filtered the gLAB data to compare to ERA5 using a six-hour Gaussian window and then matched at the one-hour time steps.  This procedure matched the filtered gLAB data's fourier transform to have approximately the cutoff frequency of ERA5. This comparison for one week is plotted in Figure~\ref{fig:rubinPWVsmooth}. A scatter plot of ERA5 and GNSS PWV for all the data is presented in Figure~\ref{fig:PWVscatter}.  We measure a Pearson correlation coefficient of \RgLABERA{} indicating strong correlation between the two datasets.  We find the mean difference between our smoothed gLAB results and ERA5 to be \biasgLABERA~\PWVunits{} and an RMSE of \PWVRMSEgLABERA~\PWVunits{}.
Several authors have compared pressure and temperature from ERA5 and meteorological sensors \citep[see, e.g.,][]{Ssenyunzi20, Wang19}.  The biases they measure in pressure $\lesssim 1.6$~hPa translate to a 0.5~\PWVunits{} bias in the resulting PWV, which would be consistent with our mean difference.  There is also evidence that ERA5 is less accurate in mountainous regions.  For instance, \citet{Wei24} find that the relative error in relative humidity can range from $40\%-100\%$ in an altitude range of 500~m to 2500~m. However, the goal of this comparison is not to evaluate the accuracy of ERA5, but rather to demonstrate that we get excellent agreement in bulk trends doing a relatively simple comparison.

\begin{figure}
  \centering
  \includegraphics[width=0.8\textwidth]{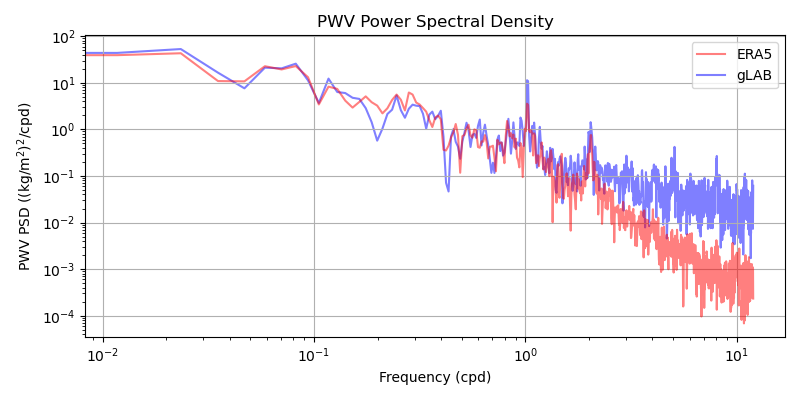}\\
  \caption{PWV power spectral density (PSD) using Welch's method for ERA5 (red) and GNSS (blue).  The x-axis is in cycles per day and the y-axis is (\PWVunits{})$^2$/cycles per day.  The general shape of both PSDs is similar, with significant power at 1 day and 12 hours.}
  \label{fig:PSD}
\end{figure}

We compute the PWV power spectral density (PSD) for ERA5 and gLAB using Welch's method in Figure~\ref{fig:PSD}. We first used a FFT to show that the ZTD screening removes orbital artifacts (Section \ref{sec:processingstrategy}). For the comparison with ERA5 we use Welch's method, whose averaging over overlapping windowed segments reduces the variance of the spectral estimate compared to a single FFT of a series that now contains more interpolated gaps. The spectra behave similarly between gLAB and ERA5 with clear signals at approximatey 1 day and 12 hours. The gLAB data in this figure is not smoothed, which explains the higher noise plateau. There are slight peaks in ERA5 at harmonics of 1 day, which implies that not all of the signal in gLAB at daily harmonics can be a result of orbital artifacts. Both of these facts indicate that there is some subdaily structure to PWV.

\begin{figure}
  \centering
  \includegraphics[width=0.8\textwidth]{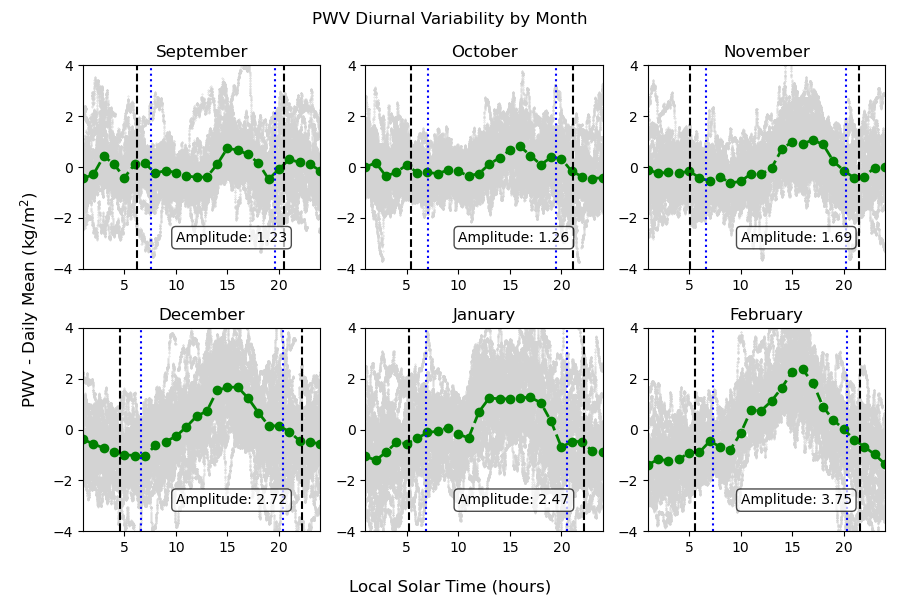}\\
  \caption{Daily variations of PWV as a function of Local Solar Time according to month.  The diurnal variation is calculated as PWV minus the daily mean.  Plotted in light gray are all the measurements. The dotted green line is the average for each hour. There is a clear daily signal as we move into warmer months. February has the highest variation at \FebruaryAmplitude{}~\PWVunits{}. Approximate locations for astronomical twilight ($18^{\circ}$) are marked as the black-dashed line. The blue-dotted lines indicate approximate sunrise and sunset times. There is a noticeable daily decrease in PWV after sunset from November to February.}
  \label{fig:diurnalPWV}
\end{figure}

To further investigate the diurnal variability of PWV we subtract the daily mean and plot the difference in Figure~\ref{fig:diurnalPWV}.  The data are segmented by month. The data in this figure have not been smoothed.  On the x-axis we transform the data from UTC to Local Solar Time (LST). This system aligns noon with the time the Sun is directly overhead. The y-axis is in \PWVunits{}. The black-dashed lines indicate approximate astronomical twilight and the blue-dotted lines indicate approximate sunrise and sunset times.

A clear daily signal emerges as we transition from spring to summer. Starting in November and continuing through February, the PWV begins to increase around 11~LST and then plateaus for several hours around 16~LST before decreasing again in the early evening. The variation is greatest in February with an average amplitude of \FebruaryAmplitude~\PWVunits{}. The daily variations are around 1~\PWVunits{} in September and October.  The average PWV amplitude over the observing night (evening to morning astronomical twilight) ranges from 0.4~\PWVunits{} to 0.9~\PWVunits{}.

PWV is affected by the global weather pattern and local transient weather systems.  At Rubin Observatory, PWV is also modulated by orography and land-sea interactions.  Broadly speaking, several hours after sunrise, heating of the Andes slopes produces anabatic winds from the west driving moisture uphill.  During the night and early morning hours, a weaker eastward wind develops and carries moisture down the mountain \citep{Rutllant03, Vicencio24, Schween20}.

\begin{figure}
  \centering
  \includegraphics[width=0.7\textwidth]{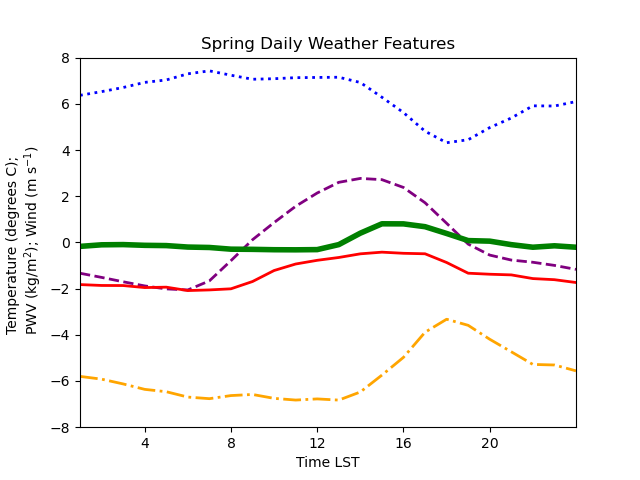}\\
    \includegraphics[width=0.7\textwidth]{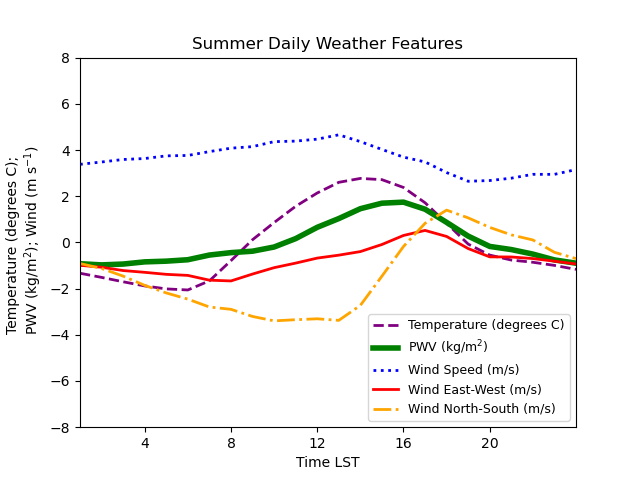}\\
  \caption{Daily weather fluctuations at Rubin Observatory in spring (top) and summer (bottom). Variations in temperature (purple-dashed) and PWV (bold-green) are calculated as the difference from the daily mean. Wind speed (blue-dotted), and wind north-south (orange-dot-dashed) and east-west (solid-red) components are simply averaged over each hour. Negative values in this case indicate wind is flowing south or west. In spring there is a moderate correlation between PWV and east-west wind component.  The correlation is higher in summer with $R = \summerRwind$.  Heating of the nearby Cachacos Valley produces winds coming from the southwest which summit in summertime and may be a mechanism to transport water vapor. }
  \label{fig:dailyweather}
\end{figure}

In Figure \ref{fig:dailyweather} we compare the daily variations in PWV  along with fluctuations in temperature in $^{\circ}$C, and the east-west and north-south wind components.  At a latitude of ~$\approx 30^{\circ}$S and at an altitude of 4~km, there is a large-scale west flowing wind.  Below this, winds hit the west slope of the Andes and are redirected to form a permanent northerly flow \citep{Kalthoff02}. Thus we observe a predominant wind coming from the North at Rubin Observatory in spring time. An easterly wind tends to blow to the west and we have moderate correlation between the east-west wind and PWV with $R = \springRwind$.

We observe that PWV is more correlated with east-west wind direction in summer, with $R = \summerRwind$. As we move into summer, heating of the nearby Cachacos Valley produces an afternoon southwest valley wind at CTIO and Rubin Observatories \citep{Kalthoff02}. The daily PWV signal could result from the transport of moisture along these valley winds.  In the spring these winds may only weakly reach the summit, accounting for the decreased daily PWV signal.   In the evening a dry, katabatic eastward wind develops and PWV remains constant.  The terrain at Rubin Observatory is very complex and the true underlying processes are much more involved than the simple picture we present here.  Nevertheless, measuring the daily structure of PWV and its correlation with east-west wind motivates further investigation into measuring north and east troposphere gradients with GNSS.

\subsection{Implications for LSST Calibration}

Our GNSS-PWV system provides a continuous, absolute measurement of the isotropic (zenith-equivalent) column of precipitable water vapor above Rubin Observatory, independent of sky conditions, time of day, or telescope pointing. For a given exposure, this can be projected onto the line of sight with an airmass scaling, which is appropriate if PWV is approximately homogeneous across the sky on the relevant timescales. \citet{Li18} tested this at Apache Point Observatory using high-resolution, near-infrared calibration spectra and found that PWV is homogeneous across the sky at a given epoch for most observations. APO is at a similar altitude, but different climate than Cerro Pachón. GNSS troposphere gradient estimates offer one way to quantify departures from isotropy at our site.

The GNSS system complements line-of-sight spectroscopic monitoring such as AuxTel \citep{LATISS, Neveu24}, which measures water absorption but also captures aerosol and other atmospheric transmission parameters. The GNSS receiver provides an absolute, continuous water-vapor-only constraint at lower operational cost. Together they could cross-validate each other, and a GNSS-based absolute scale could help calibrate AuxTel-derived PWV.

At present, our GNSS-PWV measurements are not incorporated into the LSST calibration analysis. Once per-exposure FGCM PWV estimates are available, they can be compared directly with the GNSS-derived values, and any differences can be used with atmospheric transmission templates to correct the photometry. For high-precision applications such as supernova cosmology, one could multiply a model supernova spectrum with a PWV transmission function derived from the GNSS measurements to compute a per-epoch, per-band flux correction accounting for the chromatic PWV absorption. We recommend that GNSS-PWV measurements be incorporated into the LSST calibration pipeline.

\section{Data Access}\label{sec:dataaccess}

The main data product of this work is presented in Table~\ref{table:rubin_pwv}.
The full data file is available in the online version of the article.
Upon publication the code necessary to replicate these results will be available at \dataset[doi:10.5281/zenodo.24483677]{https://doi.org/10.5281/zenodo.24483677}.

\begin{itemize}
  \item Table~\ref{table:rubin_pwv} presents the gLAB-calculated ZTD and error, Rubin weather data, and resulting PWV and error.
  \item Scripts to run gLAB with the configuration outlined in Section \ref{sec:processingstrategy}.
  \item Code and notebooks to generate all the figures and statistical analysis in this paper.
\end{itemize}

The raw GNSS RINEX observation and navigation files as recorded by the receiver total 2~GB and are available upon request to either AW or MWV.

\begin{table}
\caption{Rubin Observatory PWV, September 2025 -- February 2026}
\label{table:rubin_pwv}
\begin{tabular}{lrrrrrr}
\toprule
Date Time & ZTD & $\sigma_{\text{ZTD}}$ & Temperature & Pressure & PWV & $\sigma_{\text{PWV}}$ \\
& (m) & (m) & ($^\circ$C) & (Pa) & (\PWVunits{}) & (\PWVunits{}) \\
\tableline
2025-09-05 00:00:30 & 1.7028 & 0.0027 & 5.0 & 73953.6 & 2.41 & 0.42 \\
2025-09-05 00:00:45 & 1.7028 & 0.0027 & 5.0 & 73953.6 & 2.41 & 0.42 \\
2025-09-05 00:01:00 & 1.7030 & 0.0027 & 5.0 & 73953.6 & 2.44 & 0.42 \\
2025-09-05 00:01:15 & 1.7031 & 0.0027 & 5.0 & 73938.6 & 2.51 & 0.42 \\
2025-09-05 00:01:30 & 1.7032 & 0.0027 & 5.0 & 73938.6 & 2.52 & 0.42 \\
2025-09-05 00:01:45 & 1.7036 & 0.0026 & 5.0 & 73938.6 & 2.59 & 0.41 \\
2025-09-05 00:02:00 & 1.7036 & 0.0026 & 5.0 & 73953.6 & 2.53 & 0.41 \\
2025-09-05 00:02:15 & 1.7037 & 0.0026 & 5.0 & 73953.6 & 2.55 & 0.41 \\
2025-09-05 00:02:30 & 1.7038 & 0.0026 & 5.0 & 73953.6 & 2.56 & 0.41 \\
\tableline
\end{tabular}
\vspace{10pt} \\
Notes: ZTD and formal error as determined by gLAB in PPP mode.  Temperature as reported by the Rubin Weather Station which is not at equal altitude with the GNSS receiver. An altitude adjustment to pressure is applied assuming the pressure sensor is 7~m below the receiver. PWV is calculated using the Saastamoinen model for ZHD and the Bevis $T_m$ relationship to convert ZWD to PWV.  A full version of this table can be found in the online article resources.
\end{table}

\section{Future Work}\label{sec:futurework}

This study provides insights into PWV at Rubin Observatory and outlines an easy-to-implement GNSS processing strategy that can by employed at other observatories to measure PWV.
Future work should focus on addressing some of the limitations of the current analysis with gLAB software.
Of highest priority will be looking for robust indicators of PWV variation across the sky.
Measuring the first order asymmetry of the troposphere may prove important at a location like Rubin Observatory where there are prevailing winds and daily water vapor transport cycles.
For instance, other PPP analysis packages provide estimate of the north and east troposphere gradients in addition to ZTD and position.
Other analysis choices, specifically regarding the choice of mapping function, higher order ionosphere corrections, and ambiguity fixing should also be investigated.

The conversion between ZTD and PWV could also be a source of systematic error and should be examined in more detail.  While the Saastamoinen model for the hydrostatic delay and the Bevis approximation for $T_m$ are considered to be accurate for many locations, their validity at Rubin Observatory could be verified with, e.g., radiosonde measurements.

Long-term future work will focus on line-of-sight PWV measurements appropriate for each observed LSST image.
An important step will be to utilize additional GNSS constellations like Galileo, GLONASS, and BeiDou.
Understanding inter-system calibration biases between the different constellations will be vital.
Current precise clock analysis files are processed in batches.
As a result, day boundary discontinuities exist at the boundary epochs of adjacent processing runs.
IGS final clock products are the most consistent, with discontinuities less than 0.1~ns; however, when we used other analysis center products (not presented) noticeable signals at the boundaries were present. Resolving discontinuities in orbit and clock files, and understanding inter-system calibration biases, will both be important for continuous multi-GNSS constellation PWV monitoring.

Rubin Observatory has an auxiliary calibration telescope (AuxTel) with a spectrograph that is observing stars during the night.  These observations are taken every $\sim$5 minutes and provide very detailed information on the absorption structure along the specific line-of-sight to the star observed.
A detailed comparison to the AuxTel observations of stars to the all-sky, PWV measurements will provide robustness and verification of both approaches to calibration of the atmosphere above Rubin Observatory.

\section{Summary}\label{sec:summary}

In this paper we present 6 months of continuous measurements of PWV at Rubin Observatory using data from a dual-band GNSS receiver processed with the gLAB software.
We found gLAB to be an easy to use software package and the supporting documentation and tutorials to be excellent stepping stones into the intricacies of GNSS processing.

First, we validated the software and our processing strategy against the IGS position and troposphere products.
We compared position and ZTD measurements against 10 IGS stations that utilize the same Trimble GNSS receiver.
We found that gLAB can recover receiver position at the mm level in each direction using only the GPS constellation.
The RMSE between gLAB and IGS measurements is 9~mm in ZTD.
This is an improvement over previous results likely due to the updated version of gLAB and the longer convergence time we employ.
We find, conservatively, that our setup achieves a ZTD accuracy of 6~mm, which translates to a PWV accuracy of \PWVaccuracy~\PWVunits{}.

We present one month of commissioning data from Allegheny Observatory, located near Pittsburgh, PA.
PWV at this observatory is dynamic and highly variable ranging from less than 5~\PWVunits{} to over 40~\PWVunits{}.
We compare our measurements to the ERA5 weather model hourly total column water vapor measurement.
After adjusting for the altitude difference between ERA5 and the GPS receiver, we find excellent agreement with a mean PWV difference of 0.2~\PWVunits{} and a standard deviation of 1.4~\PWVunits{}.
This comparison is not entirely independent, as we also utilize the ERA5 temperature and pressure measurements as part of our calculation to estimate PWV.

We deployed a GNSS receiver an antenna at Rubin Observatory in 2025 September 4 and have been recording data continuously since. We emphasize that meteorological monitoring is required to convert GNSS data to PWV.
We measure PWV with \PWVprecision{}~\PWVunits{} precision.
Results from spring (September, October, November) monitoring show average PWV values of \springPWVmean{}~\PWVunits{} and daily fluctuations of
\SeptemberAmplitude{}~\PWVunits{} increasing to \NovemberAmplitude{}~\PWVunits{} by the end of spring.
PWV continues to increase into the summer season with an average value of \summerPWVmean{}~\PWVunits{} and daily fluctuations increasing to \FebruaryAmplitude~\PWVunits{}.
We compare our measurements to the ERA5 weather model using hourly data on pressure levels and find good correlation of $R = \RgLABERA{}$ and an overall bias of \biasgLABERA{}~\PWVunits{}.

Finally, we examine the daily variations of PWV at Rubin Observatory.
Water vapor begins to increase daily around 11~LST (local {\em solar} time) until about 16~LST.
It then continues to fall over the early evening and maintains a fairly constant value from midnight until late morning.
The average amplitude of this signal is \meanSpringAmplitude{}~\PWVunits{} in spring and \meanSummerAmplitude{}~\PWVunits{} in summer.
We find the diurnal variation of PWV is highly correlated with the east-west wind component.
We hypothesize that this is due to moisture from the west traveling east due to anabatic winds caused by daily heating of the Cachacos Valley.
These results motivate further investigation into the asymmetry of the troposphere at the Vera C. Rubin Observatory to reach the full potential of calibration of atmospheric absorption due to PWV.

\section{acknowledgements}
This work is supported in part by the M. Hildred Blewett Fellowship of the American Physical Society, www.aps.org. MWV was supported in part by US DOE Office of Science grant DE-SC0007914. This material is based upon work supported in part by the National Science Foundation through Cooperative Agreements AST-1258333 and AST-2241526 and Cooperative Support Agreements AST-1202910 and AST2211468 managed by the Association of Universities for Research in Astronomy (AURA). This work is also funded in part by the NASA Pennsylvania Space Grant Consortium.

This paper has been through and approved by the Rubin Project Publication Review process. We thank Eli Rykoff, who served as the Rubin/LSST internal reviewer for helpful questions that improved the analysis and validation of results in this paper.
We thank and acknowledge the contributions of
Daniel Perrefort for work on early conceptual design and discussions of Rubin installation.
We thank Christopher W. Stubbs for support of the idea and assistance in acquiring GNSS hardware.
We thank Brian Stalder for help with obtaining data from the Rubin weather monitoring system and details of the Rubin site topography.
We thank members of the Rubin Construction Team, in particular Mario Rivera for physical construction of the mount and installation of antenna and receiver; and Fritz Muller for connecting the receiver to power and internet and establishing it on the summit.

A. Weyant led the data analysis, IGS validation, and paper writing. A. Weyant also developed the GNSS processing methodology including investigation and evaluation of gLAB GNSS processing software and atmospheric modelling. MWV served as PI and was involved in planning, logistics, data archiving, receiver configuration, analysis and manuscript review. E. Urbach oversaw coordination and project management of observatory operations and provided consulting on data analysis. S. Gelven contributed to the Allegheny Observatory and ERA5 data analysis and processing.  M. Tamargo aided in gLAB exploration and commissioning of receiver at Allegheny Observatory.

The data used in this study were acquired as part of NASA's Earth Science Data Systems and archived and distributed by the Crustal Dynamics Data Information System (CDDIS).

Copernicus programme: Generated using or contains modified Copernicus Climate Change Service information . Neither the European Commission nor ECMWF is responsible for any use that may be made of the Copernicus information or data it contains.

We used Claude Sonnet 5.5 (Anthropic) to discuss the error contributions (interactive chat) in Section~\ref{sec:pwvuncertainty}, to suggest literature, to review the glossary of terms, and to suggest edits and wording for individual passages.  Equations and citations were verified by the authors against the primary sources.  The full manuscript was not processed by an AI tool. The analysis code was written by the authors. The final analysis notebooks were screened for bugs by Claude, which also wrote the markdown summary at their top. The authors reviewed all AI-assisted content and take responsibility for the contents of the paper.

\software{gLAB \citep[v6.0.0;][]{gLAB} (\url{https://gage.upc.edu/en/learning-materials/software-tools/glab-tool-suite})}

\appendix
\section{Dictionary of Terminology for GNSS Analysis}\label{sec:appendix}
This paper uses ideas and terminology from GNSS and astmospheric modeling.  For the convenience of the astronomer we provide the following dictionary of terms.
\subsection{Observables and Signal Processing}
\begin{description}
  \item[Carrier Phase]The phase of the sinusoidal carrier wave emitted by a
    GNSS satellite, as measured by the receiver. It is very precise
    (mm-level), but ambiguous because the number of whole wavelengths between
    the satellite and receiver is unknown. Incorporating carrier phase
    measurements into the navigation equations is key to achieving mm-level
    positioning precision in PPP.
  \item[Carrier Phase Ambiguities]The unknown integer number of wavelengths
    between the satellite and receiver antenna at the start of tracking. The
    ambiguity is constant over a continuous tracking arc. gLAB does not
    attempt integer ambiguity resolution; instead it estimates each ambiguity
    as a real-valued (float) parameter in the Kalman filter.
  \item[Cycle Slip]  A discontinuity in the tracked carrier phase, caused when
    the receiver temporarily loses contact with the satellite and loses count
    of the wave cycles. After a cycle slip the ambiguity changes, so a new
    float ambiguity must be estimated for that satellite.
  \item[Global Navifation Satellite Systems (GNSS)] An overarching term for
    satellite constellations that provide geospatial position and include GPS,
    GLONASS, Galileo and others.
  \item[Ionosphere-free combination] The linear combination of signals at two different
    frequencies that removes 99.9\% of the ionopshere. This is possible because the
    ionospheric delay is dispersive and scales inversely with the frequency squared.
    The cost is that the combination amplifies measurement noise.
  \item[Phase Center Offset (PCO) and Phase Center Variations (PCV)] PCOs are the systematic offsets between an antenna's physical reference point and its mean electical phase center for a given frequency. The PCVs account for deviations as a function of incoming signal elevation and azimuth. Failing to model these results in systematic offsets, particularly in estimate height and ZTD. Their corrections are contained in the ANTEX file.
  \item[Precise Point Positioning] A technique that utilizes carrier and
    code signals from a single receiver and can achieve mm- to cm-level
    precision in positioning. It requires knowledge of the precise orbits and
    clock parameters of the satellites, so with final products cannot be performed in real time.
    The zenith tropospheric delay is estimated alongside
    the position, which is what makes PPP useful for PWV retrieval.
  \item[Pseudorange] The apparent range to the satellite and is the basic observable in GNSS positioning.  It is not equal to the geometric range due to effects such as delays from the atmosphere and synchronisation errors between the receiver and satellite clocks.
  \item[Ranging Code] Also called the Pseudo-Random Noise (PRN) code, is a unique binary signal emited continuously by a GNSS satellite and is used to measure travel time of the radio signal between a satellite and antenna. The
    receiver finds this travel time by correlating the incoming signal with a
    replica of the code. This travel time multiplied by the speed of light gives us the pseudorange.
\end{description}
\subsection{Products and File Formats}
\begin{description}
  \item[ANTEX] This file contains the antenna phase center calibrations for satellite and receiver antennas. It can be found at \url{https://igs.org/wg/antenna/\#files}.
  \item[Crustal Dynamics Data Information System (CDDIS)] Maintains an archive of IGS products, which include precise orbit and clock files. These can be found at \url{https://cddis.nasa.gov/archive/gnss/products/}.
  \item[Internation GNSS Service (IGS)] An organization of research universities, institutions, and self funding agencies that work together to provide open access to high quality GNSS products that support a variety of applications. Learn more about IGS products at \url{https://igs.org/products/}.
  \item[Orbit (SP3) and Clock (CLK) products] IGS Analysis Centers provide GNSS satellie orbit and clock solutions, which are combined to form the official IGS products.  The \emph{final} orbits and clocks products are the most consistent and highest quality and are available after approximately three weeks.
  \item[Receiver Independent Exchange Format (RINEX)] Standard data format to store raw GNSS data. Important types include observation files (.o) which contain the pseudo-range and carrier-phase information, and the navigation files (.n, .g) which contain satellite ephemeris, orbit, and clock parameters. These extensions follow the RINEX 2 convention.
    \item[Solution INdependent EXchange format (SINEX)] : a standard text format
used by the International GNSS Service (IGS) and the geodetic community to
exchange the results of GNSS analyses, including estimated station
coordinates and their covariances. We use the daily IGS SINEX solutions only
to compare the station positions determined by gLAB with the IGS values.
\end{description}
\subsection{Tropospheric Modelling}
\begin{description}
  \item[Conversion Factor ($\Pi$)] This conversion factor transforms ZWD to PWV and is derived in \citet{Bevis94}. It depends on the weighted mean temperature of the atmosphere $T_m$ and other constants.
  \item[Mapping function] Relates the delay along the line-of-sight to the zenith direction as a function of satellite elevation angle.  We use the Niell wet mapping function \citep{Niell96} in the gLAB software to project water vapor to the zenith direction.  A choice in mapping functions has been shown to impact ZTD.
  \item[Saastamoinen Model] A model used to estimate ZHD presented in \citet{Saastamoinen72}. It assumes hydrostatic equilibrium and requires an accurate estimate of the pressure at the height of the receiver.
  \item[$T_m$] The weighted mean temperature of the atmosphere.  We use the \citet{Bevis92} approximation to $T_m$ which depends on the surface temperature. This parameter is necessary to calculate $\Pi$.
  \item[Zenith Hydrostatic Delay (ZHD)] The delay in the arrival time of the signal due to dry gases in the atmosphere assuming the atmosphere is in hydrostatic equilibrium. We compute this using the Saastamoinen model which depends on surface pressure. This component is subtracted from the ZTD to obtain ZWD.
  \item[Zenith Total Delay (ZTD)] The total time delay due to the troposphere.  It is decomposed into a dry component (ZHD) and a wet component (ZWD) that is due to water vapor in the atmosphere.  This is the main quantity of interest that comes out of the gLAB software.
  \item[Zenith Wet Delay (ZWD)] The delay in the arrival time of the signal due to water vapor in the atmosphere. It is expressed in the zenith direction, projected from each satellite-receiver line-of-sight using a mapping function.  This parameterization of the troposphere is necessary to prevent rank deficiency when solving the navigation equations: the wet delay cannot be measured independently along each line-of-sight. Multiplying ZWD by the conversion factor $\Pi$ gives the PWV.
\end{description}

\bibliographystyle{aasjournalv7}
\bibliography{pwv_rubin.bib}

\end{document}